%% file: main.tex
\documentclass[conference,compsoc]{IEEEtran}

\ifCLASSOPTIONcompsoc
  \usepackage[nocompress]{cite}
\else
  \usepackage{cite}
\fi

\ifCLASSINFOpdf
\else
\fi

\usepackage{tikz}          %
\usepackage{amsmath}
\usepackage[ruled]{algorithm2e}
\usepackage{graphicx}
\usepackage{booktabs}
\usepackage{latexsym}   %
\usepackage{array}      %
\usepackage{multirow}
\usepackage[noend]{algpseudocode}
\usepackage{threeparttable}
\usepackage{paralist}   %
\usepackage{xspace}
\usepackage{color}
\usepackage{adjustbox}
\usepackage{balance}
\usepackage{wasysym}
\usepackage{rotating}   %
\usepackage{listings}
\usepackage{tabu}
\usepackage{pifont}
\usepackage{framed}
\usepackage{multirow}

\usepackage{url}            %
\usepackage{hyperref}

\usepackage{float}

\usepackage{xpatch}
\usepackage{tcolorbox}
\tcbuselibrary{breakable, skins}
\usepackage{hyperref}

\newcommand{\red}[1]{\textcolor[rgb]{1.00,0.00,0.00}{#1}}
\newcommand{\blue}[1]{\textcolor[rgb]{0.00,0.00,1.00}{#1}}
\newcommand{\green}[1]{\textcolor[rgb]{0.00,0.60,0.00}{#1}}

\definecolor{wheat1}{rgb}{1.000000,0.905882,0.729412}
\definecolor{LightGray}{rgb}{0.827451,0.827451,0.827451}

\newcolumntype{a}{>{\columncolor{wheat1}}l}

\definecolor{mygreen}{rgb}{0,0.6,0}
\definecolor{mygray}{rgb}{0.5,0.5,0.5}
\definecolor{mymauve}{rgb}{0.58,0,0.82}
\definecolor{darkblue}{rgb}{0.0,0.0,0.6}
\definecolor{maroon}{RGB}{102, 0, 0}
\definecolor{Maroon}{cmyk}{0,0.87,0.68,0.32}
\definecolor{darkred}{RGB}{139, 0, 0}
\definecolor{forestgreen}{RGB}{34, 139, 34}

\lstdefinelanguage{XML}
{
  basicstyle=\ttfamily\small,   %
  morestring=[b]",
  moredelim=[s][\color{darkblue}]{<}{\ },
  moredelim=[s][\color{darkblue}]{</}{>},
  moredelim=[l][\color{darkblue}]{/>},
  moredelim=[l][\color{darkblue}]{>},
  morecomment=[s]{<?}{?>},
  morecomment=[s]{<!--}{-->},
  stringstyle=\color{darkred},
  identifierstyle=\color{mymauve}
}

\lstdefinestyle{customJava}{
  breaklines=true,
  keepspaces=true,
  frame=single,
  language=Java,
  showstringspaces=false,
  basicstyle=\footnotesize\ttfamily,
  keywordstyle=\color{blue},
  otherkeywords={+, getIntent},
  numbers=left,
  numbersep=5pt,
  numberstyle=\scriptsize\color{black},
  rulecolor=\color{black},
  stepnumber=1,
  tabsize=2,
  commentstyle=\itshape\color{green!40!black},
  stringstyle=\color{orange},
  emph=[1]  %
  {
        do,
        try,
        new,
        catch,
        while,
        SecProvider,
        SecReceiver,
        SecService,
        SecActivity,
        SecSink,
  },
  emphstyle=[1]{\color{darkred}},
  emph=[2]  %
  {
        @Override,
  },
  emphstyle=[2]{\color{purple!40!black}},
  belowskip=-1em, %
}

\newif\ifANNOYMIZE
\ANNOYMIZEtrue

\newif\ifACM
\ACMtrue  %

\ifACM
\fi

\ifACM
\newcommand{\myfig}{Figure\xspace}
\else
\newcommand{\myfig}{Fig.\xspace}
\fi

\ifACM
\newcommand{\mysec}{\S}
\else
\newcommand{\mysec}{\S}
\fi

\newcommand{\cfr}{Code4Rena }

\newcounter{findingCounter}

\newcommand{\finding}[2]{
  \begin{tcolorbox}[enhanced, left=3mm,right=3mm,
    colback=gray!10, colframe=gray!80, boxrule=0pt,
    borderline west={4pt}{0pt}{gray!90},
    breakable
    ]
    \stepcounter{findingCounter}
    \textbf{Finding #1:} #2
    \end{tcolorbox}
}

\newcounter{knowledgeCounter}

\definecolor{cadmiumgreen}{rgb}{0.0, 0.42, 0.24}

\newcommand{\etal}{et al.\xspace}

\newcommand{\name}{LLM4Vuln\xspace} %

\newcommand{\dataset}{UniVul\xspace}
\newsavebox{\bigimage} %

\input{code/sol-highlighting}

\begin{document}

\title{\name: Extracting Zero-shot Vulnerability Knowledge and Supplying It to GPT for Self-Evolving Vulnerability Detection}
\title{Extracting Zero-shot Vulnerability Knowledge for Self-Evolving LLM-based Vulnerability Detection: Feasibility, Comparison, and Insights}
\title{\name: Extracting Hierarchical Vulnerability Knowledge for Self-Evolving LLM-based Vulnerability Detection in Smart Contracts}
\title{Turning ChatGPT into \name: An Experimental Framework for Evaluating and Enhancing LLM's Vulnerability Reasoning}
\title{\name: An Unified Framework for Evaluating, Decoupling, and Enhancing LLMs' Vulnerability Reasoning}
\title{Towards a Unified Evaluation Framework for Decoupling and Enhancing LLMs' Vulnerability Reasoning: An Empirical Study on Smart Contracts}
\title{\name: A Unified Evaluation Framework for Decoupling and Enhancing LLMs' Vulnerability Reasoning}

\IEEEoverridecommandlockouts

\author{\IEEEauthorblockN{%
  Yuqiang Sun\IEEEauthorrefmark{1}, 
  Daoyuan Wu\IEEEauthorrefmark{2}\thanks{Daoyuan Wu is the corresponding author.}, 
  Yue Xue\IEEEauthorrefmark{3}, 
  Han Liu\IEEEauthorrefmark{2}, 
  Wei Ma\IEEEauthorrefmark{4}, 
  Lyuye Zhang\IEEEauthorrefmark{1}, 
  Yang Liu\IEEEauthorrefmark{1} and 
  Yingjiu Li\IEEEauthorrefmark{5}}
\IEEEauthorblockA{\IEEEauthorrefmark{1}Nanyang Technological University, Singapore\\
                  yuqiang.sun@ntu.edu.sg, zh0004ye@e.ntu.edu.sg, yangliu@ntu.edu.sg}
\IEEEauthorblockA{\IEEEauthorrefmark{2}The Hong Kong University of Science and Technology, Hong Kong SAR, China\\
                  daoyuan@cse.ust.hk, liuhan@ust.hk}
\IEEEauthorblockA{\IEEEauthorrefmark{3}MetaTrust Labs\\
                  nerbonic@gmail.com}
\IEEEauthorblockA{\IEEEauthorrefmark{4}Singapore Management University, Singapore\\
                  weima93@gmail.com}
\IEEEauthorblockA{\IEEEauthorrefmark{5}University of Oregon, Eugene, OR, USA\\
                  yingjiul@uoregon.edu}
}

\maketitle

\input{abstract}

\input{intro}

\input{background}
\input{formal}
\input{design}

\input{implement}

\input{evaluate}

\input{discussion}

\input{related}
\input{conclude}

\bibliographystyle{IEEEtran}
\bibliography{ref}

\input{appendix}

\end{document}

%% file: code/sol-highlighting.tex
\usepackage{listings, xcolor}

\definecolor{verylightgray}{rgb}{.97,.97,.97}
\definecolor{codegreen}{rgb}{0,0.55,0}

\lstdefinelanguage{Solidity}{
	keywords=[1]{anonymous, assembly, assert, balance, break, call, callcode, case, catch, class, constant, continue, constructor, contract, debugger, default, delegatecall, delete, do, else, emit, event, experimental, export, external, false, finally, for, function, gas, if, implements, import, in, indexed, instanceof, interface, internal, is, length, library, log0, log1, log2, log3, log4, memory, modifier, new, payable, pragma, private, protected, public, pure, push, require, return, returns, revert, selfdestruct, send, solidity, storage, struct, suicide, super, switch, then, this, throw, transfer, true, try, typeof, using, value, view, while, with, addmod, ecrecover, keccak256, mulmod, ripemd160, sha256, sha3}, %
	keywordstyle=[1]\color{blue}\bfseries,
	keywords=[2]{address, bool, byte, bytes, bytes1, bytes2, bytes3, bytes4, bytes5, bytes6, bytes7, bytes8, bytes9, bytes10, bytes11, bytes12, bytes13, bytes14, bytes15, bytes16, bytes17, bytes18, bytes19, bytes20, bytes21, bytes22, bytes23, bytes24, bytes25, bytes26, bytes27, bytes28, bytes29, bytes30, bytes31, bytes32, enum, int, int8, int16, int24, int32, int40, int48, int56, int64, int72, int80, int88, int96, int104, int112, int120, int128, int136, int144, int152, int160, int168, int176, int184, int192, int200, int208, int216, int224, int232, int240, int248, int256, mapping, string, uint, uint8, uint16, uint24, uint32, uint40, uint48, uint56, uint64, uint72, uint80, uint88, uint96, uint104, uint112, uint120, uint128, uint136, uint144, uint152, uint160, uint168, uint176, uint184, uint192, uint200, uint208, uint216, uint224, uint232, uint240, uint248, uint256, var, void, ether, finney, szabo, wei, days, hours, minutes, seconds, weeks, years},	%
	keywordstyle=[2]\color{teal}\bfseries,
	keywords=[3]{block, blockhash, coinbase, difficulty, gaslimit, number, timestamp, msg, data, gas, sender, sig, value, now, tx, gasprice, origin},	%
	keywordstyle=[3]\color{violet}\bfseries,
	identifierstyle=\color{black},
	sensitive=false,
	comment=[l]{//},
	morecomment=[s]{/*}{*/},
	commentstyle=\color{codegreen}\ttfamily,
	stringstyle=\color{red}\ttfamily,
	morestring=[b]',
	morestring=[b]"
}

%% file: abstract.tex
\begin{abstract}

Large language models (LLMs) have demonstrated significant potential in various tasks, including those requiring human-level intelligence, such as vulnerability detection.
However, recent efforts to use LLMs for vulnerability detection remain preliminary, as they lack a deep understanding of whether a subject LLM's vulnerability reasoning capability stems from the model itself or from external aids such as knowledge retrieval and tooling support.

In this paper, we aim to decouple LLMs' vulnerability reasoning from other capabilities, such as vulnerability knowledge adoption, context information retrieval, and advanced prompt schemes.
We introduce \name, a unified evaluation framework that separates and assesses LLMs' vulnerability reasoning capabilities and examines improvements when combined with other enhancements.

To support this evaluation, we construct \dataset, the first benchmark that provides retrievable knowledge and context-supplementable code across three representative programming languages: Solidity, Java, and C/C++.
Using \name and \dataset, we test six representative LLMs (GPT-4.1, Phi-3, Llama-3, o4-mini, DeepSeek-R1, and QwQ-32B) for 147 ground-truth vulnerabilities and 147 non-vulnerable cases in 3,528 controlled scenarios. %
Our findings reveal the varying impacts of knowledge enhancement, context supplementation, and prompt schemes. %
We also identify 14 zero-day vulnerabilities in four pilot bug bounty programs, resulting in \$3,576 in bounties.

\end{abstract}

%% file: intro.tex
\section{Introduction}
\label{sec:intro}

In the rapidly evolving landscape of computer security, Large Language Models (LLMs) have significantly transformed our approach to complex challenges.
Equipped with extensive pre-training and strong instruction-following capabilities, these models excel in understanding and interpreting %
the semantics of both human and programming languages.
This has led to the emergence of \textit{LLM-based vulnerability detection}\footnote{In this paper, we do not explicitly distinguish between ``vulnerability detection'' and ``vulnerability reasoning'' for LLMs, and use both terms interchangeably.}, offering superior intelligence and flexibility compared to traditional program analysis-based techniques (e.g.,~\cite{Wang_Wei_Gu_Zou_2010, Fioraldi_Maier_Eissfeldt_Heuse_2020, Cristian_klee, Shoshitaishvili_Wang_Salls_Stephens_Polino_Dutcher_Grosen_Feng_Hauser_Kruegel_et_al._2016,Arzt_Rasthofer_Fritz_Bodden_Bartel_Klein_Le_Traon_Octeau_McDaniel_2014,Wu_Gao_Deng_Rocky_2021,fang_modifier_issta_2023,YiNDSS2023}) and neural network-based detectors (e.g.,~\cite{li2018vuldeepecker,Zhou_Liu_Siow_Du_Liu_2019,Chakraborty_Krishna_Ding_Ray_2022,Chen_Ding_Alowain_Chen_Wagner_2023,Xu_Liu_Feng_Yin_Song_Song_2017,she2019neuzz}).

Under this emerging paradigm, triggered by the successful release of ChatGPT~\cite{OpenAI, ouyang_training_2022, kojima2022large} on 30 November 2022, related research primarily focuses on two dimensions.
One dimension involves designing \textit{specific LLM-based detectors} for different security problems.
For example, researchers have developed TitanFuzz~\cite{Deng_Xia_Peng_Yang_Zhang_2023}, FuzzGPT~\cite{Deng_Xia_Yang_Zhang_Yang_Zhang_2023}, Fuzz4All~\cite{Xia_Paltenghi_Tian_Pradel_Zhang_2024}, and ChatAFL~\cite{Meng_Mirchev_Bohme_Roychoudhury} for fuzzing various vulnerabilities; GPTScan~\cite{Sun_Wu_Xue_Liu_Wang_Xu_Xie_Liu_2023} and GPTLens~\cite{Hu_Huang_Ilhan_Tekin_Liu_2023} for detecting smart contract vulnerabilities; and LLift~\cite{Li_Hao_Zhai_Qian_2023} and LATTE~\cite{Liu_Sun_Zheng_Feng_Qin_Wang_Li_Sun_2023} for LLM-enhanced program and binary analysis.

The other dimension aims to \textit{benchmark or evaluate} LLMs' capabilities in vulnerability detection.
It examines how different models, configurations, and instructions influence detection results, addressing the key question, ``How far have we come?'' in LLM-based vulnerability detection.
Notably, Thapa \etal~\cite{Thapa_Jang_Ahmed_Camtepe_Pieprzyk_Nepal_2022} pioneered this effort by benchmarking transformer-based language models against RNN-based models for software vulnerability detection.
Following the release of ChatGPT and GPT-4, additional LLM-focused benchmark studies have been conducted, including for smart contracts~\cite{David_Zhou_Qin_Song_Cavallaro_Gervais_2023,Chen_Su_Chen_Wang_Bi_Wang_Lin_Chen_Zheng_2023}, traditional C/C++/Java vulnerabilities~\cite{Gao_Wang_Zhou_Zhu_Zhang_2023,Khare_Dutta_Li_Solko-Breslin_Alur_Naik_2023,Ullah_Han_Pujar_Pearce_Coskun_Stringhini_2023,ding2024vulnerability,Lin2023from}, and vulnerability repair~\cite{Pearce_Tan_Ahmad_Karri_Dolan-Gavitt_2023,zhang2024acfix}.

\begin{figure}[t]
    \centering
    \includegraphics[width=\linewidth]{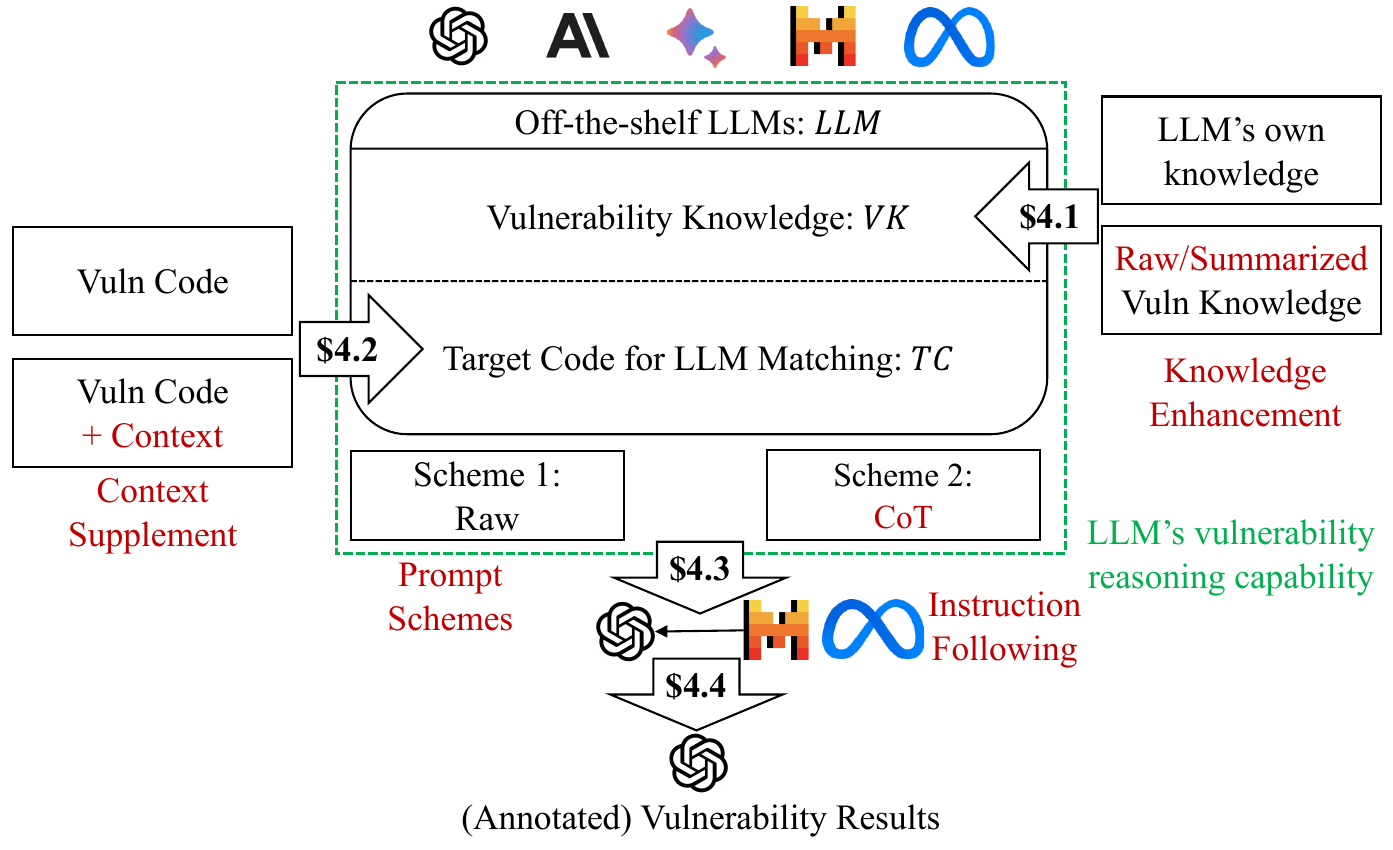}
    \vspace{-4ex}
    \caption{An illustration of the LLM-based vulnerability detection paradigm and the \name framework.}
    \label{fig:overview}
\end{figure}

Our research falls into the second dimension.
However, instead of focusing on the performance of individual LLM instances and their configurations, we delve into the paradigm itself and consider what is missing or could be improved.
To this end, we first abstract and generalize the paradigm of LLM-based vulnerability detection into an architecture shown in \myfig~\ref{fig:overview}. %
Existing LLM-based vulnerability detection typically takes a piece of target code ($TC$) and asks the $LLM$ to determine whether $TC$ is vulnerable under certain prompt schemes (e.g., role playing~\cite{Hu_Huang_Ilhan_Tekin_Liu_2023} and chain-of-thought/CoT~\cite{Wei_Wang_Schuurmans_Bosma_ichter_Xia_Chi_Le_Zhou_2022}).
However, additional information about $TC$ (e.g., the context of functions and variables involved in $TC$), which can be obtained by LLMs through invoking tool support, is often overlooked in the paradigm.

More importantly, LLMs are pre-trained up to a certain cutoff date\footnote{For example, GPT-4.1 was pre-trained using data up to June 2024, while Llama-3-8b was trained with data up to March 2023; see \mysec\ref{sec:background}.}, making it challenging for LLMs to adapt to the latest vulnerability knowledge.
In other words, it is essential to incorporate relevant vulnerability knowledge ($VK$) into the paradigm.
Furthermore, open-source LLMs typically have weaker instruction-following capabilities than OpenAI models, due to the latter being aligned with extensive reinforcement learning from human feedback (RLHF)~\cite{lambert2022illustrating}, which could indirectly affect the $LLM$'s reasoning outcome in automatic evaluation.
As we see from above, the $LLM$'s vulnerability reasoning capability can be influenced by various factors beyond the model itself and its configuration.

Based on this intuition, rather than treating LLM-based vulnerability detection as a whole for evaluation, we \textit{decouple} LLMs' vulnerability reasoning capability from their other capabilities \textit{for the first time} and assess how LLMs' vulnerability reasoning could be enhanced when combined with the enhancement of other capabilities.
As for benchmarking and tuning the model itself under different configurations (e.g., different temperatures), we defer to another dimension of related work focused on how to pre-train vulnerability-specific or security-oriented LLMs~\cite{Chen_Lacomis_Schwartz_Goues_Neubig_Vasilescu_2022, Lacomis_Yin_Schwartz_Allamanis_Le_Goues_Neubig_Vasilescu_2019, len_index_count, Li_Qu_Yin_2021, Ding_Steenhoek_Pei_Kaiser_Le_Ray_2023, Zhu_Wang_Zhou_Wang_Sha_Gao_Zhang_2023, Jiang_Wang_Liu_Xu_Tan_Zhang_2023, Gai_Zhou_Qin_Song_Gervais_2023, Guthula_Battula_Beltiukov_Guo_Gupta_2023, Pei_Li_Jin_Liu_Geng_Cavallaro_Yang_Jana_2023, SmartInv24}, which goes beyond merely being a language model.

To achieve this research objective, we propose \name, a \textit{unified evaluation framework} for decoupling and enhancing LLMs' vulnerability reasoning. %
As illustrated in \myfig~\ref{fig:overview}, \name first considers the state-of-the-art LLMs' ability to actively invoke tools for seeking additional information about $TC$, such as through function calling in proprietary OpenAI models~\cite{noauthor_function_2024} and in fine-tuned open-source models~\cite{Llama-3-8b-instruct, Phi-3-mini-128k-instruct}.
Besides supplying additional information for $TC$ through LLM-invoked tools, \name also decouples and enhances LLMs' $VK$ by providing a searchable vector database of vulnerability knowledge, similar to  retrieval-augmented generation (RAG)~\cite{Lewis_Perez_Piktus_Petroni} technique for NLP-domain knowledge enhancement.
Furthermore, \name incorporates typical prompt engineering enhancements by exploring different prompt schemes and employs the most capable GPT-4.1 model to refine the raw, unstructured output of models with less proficient instruction-following capabilities.

More specifically, we formalize the LLM-based detection paradigm (\mysec\ref{sec:formal}) and isolate three key enhancement dimensions(\mysec\ref{sec:design}): (i) vulnerability knowledge enhancement, (ii) program context supplementation, and (iii) prompt scheme optimization.
\name supports systematic exploration across these dimensions and standardizes result evaluation using a GPT-4.1-assisted annotation pipeline (\mysec\ref{sec:annotation}).

To support our evaluation, we construct \dataset, a new benchmark that provides retrievable knowledge and context-supplementable code, in \mysec\ref{sec:implement}.
\dataset spans three widely-used programming languages, Solidity, Java, and C/C++, representing both smart contract and traditional software domains.
For each language, \dataset includes both (i) a curated \textit{vulnerability knowledge base} for retrieval-based knowledge enhancement and (ii) a \textit{ground-truth code corpus} for evaluation.
Specifically, the knowledge base comprises 1,013 high-risk vulnerability reports for Solidity, and structured summaries for 77 CWE categories in Java and 86 in C/C++, each extended with 10 synthetic vulnerability reports formatted consistently with Solidity's.
The code corpus contains 51 vulnerable and 51 non-vulnerable Solidity functions, 46 vulnerable and 46 non-vulnerable Java functions, and 50 vulnerable and 50 non-vulnerable C/C++ functions.

By applying \name to \dataset for testing these 294 pieces of code across 3,528 scenarios, which include three types of knowledge enhancement (one with only LLMs' pre-trained knowledge), two context supplementation options (with or without context supplementation), and two prompt schemes (raw and CoT), and using six representative LLMs, with three traditional foundation models (GPT-4.1, Phi-3, and Llama-3) and three deep reasoning models (o4-mini, DeepSeek-R1, and QwQ-32B), we identify the following four findings (more details in \mysec\ref{sec:evaluate}): %

\begin{itemize}
    \item \textit{Knowledge Enhancement:} Vulnerability knowledge augmentation yields heterogeneous effects. For traditional foundation models, knowledge enhancement substantially improves performance in Solidity---a language with business-logic-heavy vulnerabilities---with F1 scores nearly doubling on average.
        However, for Java and C/C++, the gains are limited or even negative, likely due to LLMs' stronger pre-training on CWE-like patterns in these languages.
        Surprisingly, deep reasoning models show less benefit from external knowledge, suggesting that they may internally model vulnerability semantics more robustly.

    \item \textit{Context Supplementation:} Supplying local program context yields inconsistent improvements. While traditional foundation models tend to achieve higher peak F1 scores when context is included, deep reasoning models often perform better without it. Our results suggest that context, while occasionally helpful, can also distract LLMs if not well-aligned with the vulnerability.

    \item \textit{Prompt Schemes:} CoT prompting improves precision and reduces false positives for both model types, but its impact on recall varies by task. Deep reasoning models are more stable under CoT prompting, whereas traditional foundation models show larger variance. CoT is particularly helpful for traditional models to bridge reasoning gaps in multi-step analysis.

    \item \textit{Model Types:} Our results show fundamental differences in how traditional foundation models and deep reasoning models handle vulnerability reasoning. The former benefit more from external enhancements, while the latter demonstrate stronger out-of-the-box reasoning. %
\end{itemize}

To demonstrate real-world applicability of \name's evaluation findings, we conducted a pilot study (\mysec\ref{sec:RQ6}) by applying it to four Solidity-based projects in bug bounty programs. Using only the top-performing configuration identified by \name, we submitted 29 issues, of which 14 were confirmed as true vulnerabilities and rewarded with \$3,576 in bounties. This validates \name's practical utility in uncovering zero-day vulnerabilities.

\noindent
\textbf{Contributions.} In sum, our contributions are as follows:
\begin{itemize}
    \item We propose \name, a novel framework that decouples LLMs' vulnerability reasoning from knowledge retrieval, context awareness, and prompt design, enabling structured evaluation across these dimensions.
    \item We build \dataset, the first benchmark that provides retrievable knowledge and context-supplementable code across three representative programming languages.
    \item We conduct extensive experiments across 3,528 scenarios, producing detailed insights into the impact of knowledge, context, and prompt design on different types of LLMs.
    \item We validate \name in a real-world bug bounty setting, discovering 14 new vulnerabilities with measurable security and financial impact.
\end{itemize}

\noindent
\textbf{Artifact Availability.}
We have released the code and all raw evaluation data on an anonymous website~\cite{LLM4Vuln} for review.

%% file: background.tex
\section{Background}
\label{sec:background}

Generative Pre-training Transformer (GPT) models, exemplified by the pioneering GPT-3~\cite{ouyang_training_2022}, represent a class of extensive language models trained on diverse text corpora covering a wide range of knowledge across various domains.
As indicated in Table~\ref{tab:models}, there are different variants of such Large Language Models (LLMs), including the GPT series from OpenAI, such as GPT-3.5-turbo, GPT-4, and GPT-4.1.
There are also open-source implementations, 
Llama 2/3~\cite{Touvron_Martin_Stone_Albert_Almahairi_Babaei_Bashlykov_Batra_Bhargava_Bhosale_et} and Phi-3~\cite{abdin2024phi3technicalreporthighly}.
Llama 3, a more recent and powerful model, outperforms GPT-4-turbo in some benchmarks.
Phi-3-mini-128k is a model developed by Microsoft in 2024, which is a 3.8 billion parameter language model trained on 3.3 trillion tokens, and it could rival models such as GPT-3.5 on some benchmarks.

In addition, recent efforts in LLM development have produced \textit{deep reasoning models}, such as the o1, o3, and o4 series from OpenAI, which exhibit enhanced capabilities in tasks requiring rigorous mathematical and logical reasoning. These models are optimized not just for instruction following or general-purpose tasks, but for deeper analytical thinking.
Similarly, DeepSeek-R1~\cite{deepseekai2025deepseekr1incentivizingreasoningcapability} is an open-source model released by DeepSeek, trained using reinforcement learning techniques to incentivize step-by-step reasoning. As a result, DeepSeek-R1 shows strong performance on reasoning-intensive benchmarks and is increasingly adopted in domains requiring structured decision-making, such as program analysis.
QwQ-32B~\cite{qwq32b}, developed by Alibaba and part of the Qwen~\cite{qwen2.5} model family, offers another example of a compact yet capable reasoning model. Despite having a smaller parameter count of 32 billion, QwQ-32B demonstrates performance on par with distilled versions of DeepSeek-R1 in various reasoning benchmarks. Its efficiency–accuracy trade-off makes it a compelling option in scenarios with limited compute budgets but demanding reasoning requirements.

\begin{table}[t]
\caption{Major LLMs used for security tasks (input price and output price are in USD per 1M tokens).}
\vspace{-2ex}
\label{tab:models}
\centering
    \begin{tabular}{llrrrr}
        \toprule
    LLM                     & \begin{tabular}[c]{@{}l@{}}Max. \\Token\end{tabular} & \begin{tabular}[c]{@{}l@{}}Knowledge \\Cutoff Date\end{tabular} & \begin{tabular}[c]{@{}l@{}}Input \\Price\end{tabular} & \begin{tabular}[c]{@{}l@{}}Output \\Price\end{tabular} \\ \midrule
    GPT-4.1                       & 1M                                                   & 06/2024                                                     & 2                                                                   & 8                                                                    \\
    Phi-3-mini-128k                    & 128k                                                 & 10/2023                                                     & N/A                                                                 & N/A                                                                  \\
    Llama-3-8b                    & 8k                                                   & 03/2023                                                     & N/A                                                                 & N/A                                                                     \\ \midrule
    o4-mini                       & 200k                                                 & 05/2024                                                     & 1.1                                                                 & 4.4                                                                  \\
    QwQ-32B                       & 131k                                                 & 11/2024                                                     & N/A                                                                 & N/A                                                                  \\
    DeepSeek-R1                   & 128k                                                 & 06/2024                                                     & N/A                                                                 & N/A                                                                  \\
    \bottomrule
    \end{tabular}
\end{table}

These models can be tailored for specific applications using methods like fine-tuning or zero-shot learning, enabling them to utilize tools and address problems beyond their initial training data~\cite{Schick_Dwivedi-Yu_Dessi_Raileanu_Lomeli_Zettlemoyer_Cancedda_Scialom_2023, kojima2022large}.
OpenAI and other research groups have demonstrated the effectiveness of these approaches in preparing LLMs for interactive use with external tools.
Moreover, LLMs can engage with knowledge beyond their training datasets through skillful in-context learning prompts, even without fine-tuning~\cite{Dai_Sun_Dong_Hao_Ma_Sui_Wei_2023}.
However, not all in-context learning prompts are equally effective for tasks like vulnerability detection.
Additionally, Wei \etal introduced the ``chain-of-thought'' prompting methodology~\cite{Wei_Wang_Schuurmans_Bosma_ichter_Xia_Chi_Le_Zhou_2022}, which enhances reasoning by breaking down tasks into sequential steps.
This approach prompts LLMs to address each step individually, with each stage's output influencing the next, fostering more logical and coherent outputs.

However, the application of these techniques in vulnerability detection specifically remains an area of uncertainty.
It is unclear how these methods could be used to improve precision or recall in LLM-based vulnerability detection.
Also, the type of knowledge that could be effectively integrated into in-context learning to boost performance in vulnerability detection tasks needs further investigation and clarification.

%% file: formal.tex
\section{Problem Statement: A Formal Analysis}
\label{sec:formal}

In this section, we formalize the problem space underlying LLM-based vulnerability detection, thereby guiding the design of our unified evaluation framework, \name.
Our formalization stems from a structural abstraction of the LLM vulnerability detection paradigm (see \myfig\ref{fig:overview} in \mysec\ref{sec:intro}), which captures the interaction between multiple key components.

Let $\mathcal{L}$ denote a large language model (LLM), and $\mathcal{T}$ a piece of target code (TC) to be analyzed for potential vulnerabilities. The task is to determine whether $\mathcal{T}$ contains a vulnerability and to identify its type and cause. However, $\mathcal{L}$'s ability to do so does not solely depend on its pretrained reasoning capability.
Instead, its performance is affected by several externally enhanceable factors:

\begin{itemize}
    \item \textbf{Vulnerability Knowledge ($\mathcal{K}$)}: Supplemental knowledge retrieved externally via similarity search from vulnerability knowledge bases, representing analogous cases to facilitate in-context learning by $\mathcal{L}$.

    \item \textbf{Contextual Information ($\mathcal{C}$)}: Additional code context around $\mathcal{T}$, such as caller or related functions, which might be required to understand certain vulnerabilities.

    \item \textbf{Prompt Scheme ($\mathcal{P}$)}: The instruction or interaction pattern used to elicit reasoning from $\mathcal{L}$, such as zero-shot, CoT, or role-playing.

    \item Additionally, \textbf{Instruction-following Capability ($\mathcal{I}$)}:
        The model's ability to adhere to the output format, thereby enabling reliable and structured outputs for automatic evaluation.
\end{itemize}

Formally, the output $\mathcal{R}$ of the detection task can be viewed as a function:
\[
\mathcal{R} = f_{\mathcal{L}}(\mathcal{T}, \mathcal{K}, \mathcal{C}, \mathcal{P}, \mathcal{I})
\]
In practice, however, these components often entangle, making it difficult to isolate the contribution of $\mathcal{L}$'s inherent reasoning capability.
To address this, our goal is to \textbf{decouple} $\mathcal{L}$'s vulnerability reasoning from the auxiliary components $\mathcal{K}$, $\mathcal{C}$, $\mathcal{P}$, and $\mathcal{I}$.
This allows us to answer a fundamental question:
\emph{To what extent does an LLM's vulnerability reasoning capability originate from the model itself, and to what extent can external aids enhance an LLM's inherent vulnerability reasoning capability?}

Hence, the problem reduces to designing an evaluation framework $\mathcal{F}$ that supports:
\begin{enumerate}
    \item Measuring the baseline performance of $\mathcal{L}$ in isolation, i.e., $f_{\mathcal{L}}(\mathcal{T})$;
    \item Systematically enhancing $\mathcal{L}$ with one or more of $\mathcal{K}$, $\mathcal{C}$, $\mathcal{P}$, or $\mathcal{I}$;
    \item Quantifying the marginal utility of each enhancement, i.e., the effect of $f_{\mathcal{L}}(\mathcal{T}, \cdot)$ relative to baseline;
    \item Supporting controlled experimentation across multiple LLMs and code languages.
\end{enumerate}

This formal problem definition informs the modular design of our framework \name in \mysec\ref{sec:design}, where each component is implemented in a plug-and-play manner to enable controlled evaluation and component-wise enhancement.
Moreover, to support comprehensive empirical analysis across different LLMs, we have constructed a standardized benchmark, \dataset, as described in \mysec\ref{sec:implement}.

%% file: design.tex
\section{The \name Framework}
\label{sec:design}

\begin{figure}[t!]
    \centering
    \includegraphics[width=\linewidth]{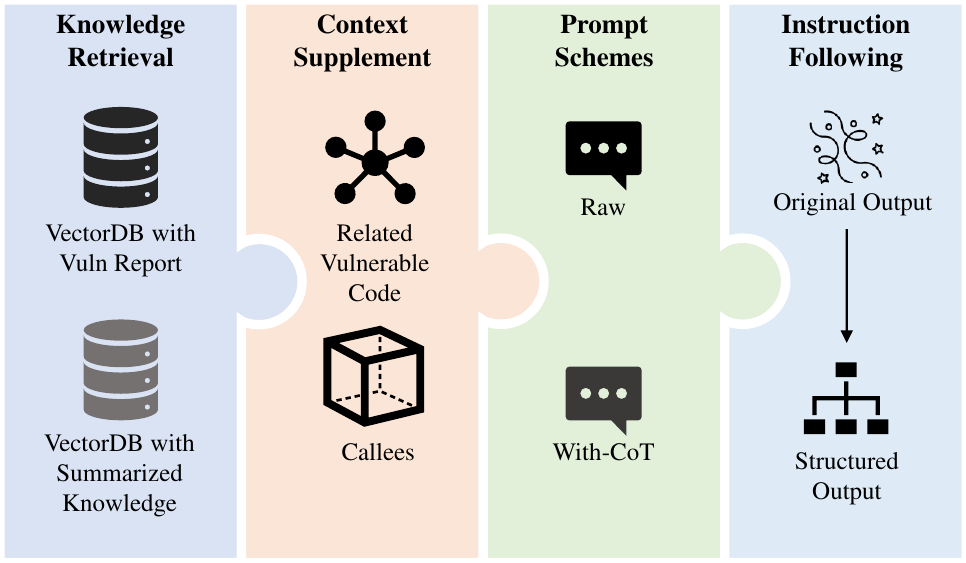}
    \vspace{-4ex}
    \caption{\name's four pluggable components for evaluating and enhancing LLMs' inherent vulnerability reasoning.}
    \label{fig:design_overview}
\end{figure}

In this section, we introduce the design of \name, a modular framework to support the paradigm of LLM-based vulnerability detection.
As illustrated in \myfig~\ref{fig:design_overview}, \name supports four types of pluggable components for evaluating and enhancing an LLMs' vulnerability reasoning capability.
These components are \textit{Knowledge Retrieval} (\mysec\ref{sec:knowledge}), \textit{Context Supplement} (\mysec\ref{sec:code}), \textit{Prompt Schemes} (\mysec\ref{sec:scheme}), and \textit{Instruction Following} (\mysec\ref{sec:scheme}).
All components are well-decoupled, allowing for easy replacement with other implementations.
For each component of \name, only one implementation is required for it to function effectively.
Lastly, for benchmarking purposes only, we design an LLM-based result annotation component in \mysec\ref{sec:annotation}.

\subsection{Vulnerability Knowledge Retrieval}
\label{sec:knowledge}

\begin{figure}[t!]
    \centering
    \includegraphics[width=\linewidth]{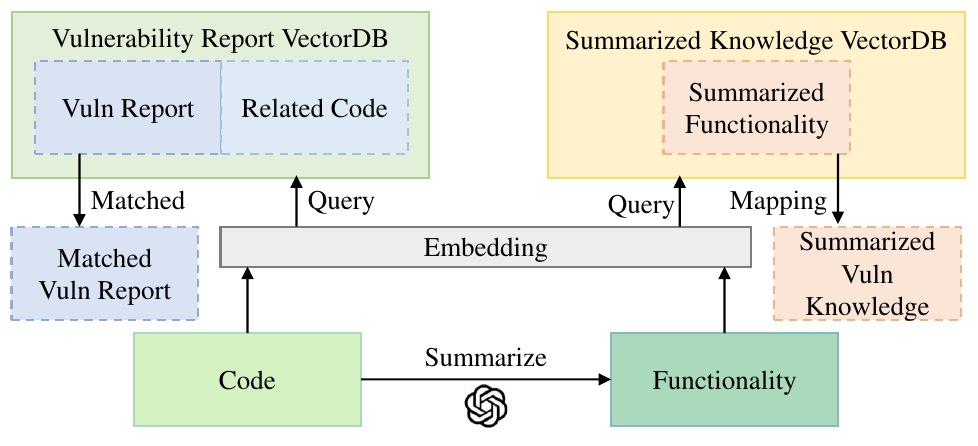}
    \vspace{-4ex}
    \caption{Two types of vulnerability knowledge retrieval.}
    \label{fig:knowledge_match}
\end{figure}

Although LLMs are trained with extensive code vulnerability data, they typically have a knowledge cutoff date for pre-training, as indicated in Table~\ref{tab:models}.
As a result, LLMs do not have up-to-date vulnerability knowledge, which is particularly crucial for detecting dynamically evolving logic vulnerabilities, such as those found in smart contracts.
To address this issue, \name proposes two types of knowledge retrieval methods for enhancing vulnerability knowledge in LLMs, as illustrated in \myfig~\ref{fig:knowledge_match}.

In the first type, as illustrated in the left part of \myfig~\ref{fig:knowledge_match}, we collect original vulnerability reports along with the corresponding vulnerable code.
We calculate their embeddings and create a vector database containing both the embeddings of the code and the associated vulnerability report.
When a target code segment $TC$ is provided, it can be used to directly search the vector database for the most similar code segments.
Since a code segment may include both code and comments, the comments can match the vulnerability reports, and the code can correspond to the code mentioned in the report.
After the retrieval process, we use only the text of the vulnerability report, excluding the code, as \textit{raw} vulnerability knowledge for subsequent analysis.

In the second type, we first use GPT-4.1 to summarize the vulnerability reports.
This summary includes the functionality of the vulnerable code and the root cause of the vulnerability, encapsulated in several key sentences.
The prompts used for summarizing the code's functionality and the key concept causing the vulnerability are listed in Appendix~\ref{sec:summarizePrompt}.
Examples of the summarized knowledge can be found in our open-source repository~\cite{FourKnowledge}.

With the functionality and knowledge from past vulnerability reports summarized, as depicted in the right part of \myfig~\ref{fig:knowledge_match}, we then calculate the embedding of this functionality part and create a vector database that contains only the functionality embeddings.
When a target code segment $TC$ is provided, we use GPT-4.1 to first summarize its functionality and then use this extracted functionality to retrieve similar functionalities in the vector database.
With the matched functionality, we can directly retrieve the corresponding vulnerability knowledge as \textit{summarized} knowledge for further analysis.

With the vector database, we can provide token-level similarity matching of knowledge to LLMs to enhance their understanding of vulnerabilities.
The proposed method can also be easily extended to include other types of knowledge in natural language.
For example, it can utilize other graph-based similarity-matching algorithms to match the control flow or data flow of the code segment with the knowledge.

\subsection{Code Context Supplementation}
\label{sec:code}

As previously illustrated in \myfig~\ref{fig:design_overview}, LLMs may detect vulnerabilities based on the provided code context of the target code $TC$.
Sometimes, the trigger of a vulnerability may be hidden in multiple functions or may require additional context to be detected.
Even if the triggering logic is entirely within the given code segment, context may help a large language model better understand the function semantics.

In this paper, we provide two types of context supplements: related vulnerable code extracted from the vulnerability reports of the vulnerable samples and the callees for all samples.
For the first type, we extract all the functions that are mentioned in the vulnerability reports, such as those from Code4Rena and issues from GitHub.
For the second type, we extract all the calling relation for all the samples, which can be used to provide a better understanding of the code.
The context supplements are processed before feeding the code into LLMs.
For benchmarking purposes only, we use static analysis to ensure the same contexts for the given code across all models, achieving a fair comparison.

In real-world LLM-based vulnerability detection, LLMs can utilize the function calling mechanism~\cite{noauthor_function_2024, Schick_Dwivedi-Yu_Dessi_Raileanu_Lomeli_Zettlemoyer_Cancedda_Scialom_2023} to assist themselves in retrieving extra context information.
For example, we can define a series of function calling APIs, such as \texttt{getFunctionDefinition}, \texttt{getClassInheritance}, \texttt{getVariableDefinition}, along with a description of the usage of each function, to assist LLMs in actively calling the function when they require the information.
More complicated contexts, such as control and data information, still need to be provided by static analysis tools though.

\subsection{Prompt Schemes and Instruction Following}
\label{sec:scheme}

In this section, we describe the enhancement of prompt schemes and instruction following.

\begin{figure}[t!]
    \begin{tcolorbox}[title=Prompt Combination: Knowledge + Output + Scheme]
        \textbf{Knowledge}\\
        \textbf{Prefix 1 - LLM's own knowledge:}\\
        As a large language model, you have been trained with extensive knowledge of vulnerabilities. Based on this past knowledge, please evaluate whether the given smart contract code is vulnerable.

        \textbf{Prefix 2 - Raw knowledge:}\\
        Now I provide you with a vulnerability report as follows: \blue{\{report\}}.
        Based on this given vulnerability report, pls evaluate whether the given code is vulnerable.

        \textbf{Prefix 3 - Summarized knowledge:}\\
        Now I provide you with a vulnerability knowledge that \blue{\{knowl\}}.
        Based on this given vulnerability knowledge, evaluate whether the given code is vulnerable.

        \tcbline
        \textbf{Output Result:}\\
        In your answer, you should at least include three parts: yes or no, type of vulnerability (answer only one most likely vulnerability type if yes), and the reason for your answer.
        
        \tcbline
        \textbf{Scheme 1 - Raw:}\\
         Note that if you need more information, please call the corresponding functions.

        \textbf{Scheme 2 - CoT:}\\
         Note that during your reasoning, you should review the given code step by step and finally determine whether it is vulnerable. For example, you can first summarize the functionality of the given code, then analyze whether there is any error that causes the vulnerability. Lastly, provide me with the result.

    \end{tcolorbox}
    \vspace{-2ex}
    \caption{Two prompt schemes combined with three different knowledge prefixes, yielding six detailed prompts.}
    \label{fig:scenarioprompt}
\end{figure}

\noindent
\textbf{Prompt Schemes.}
As described in \mysec\ref{sec:knowledge}, there are two types of knowledge provided to LLMs: the original vulnerability reports and the summarized knowledge of vulnerabilities.
Additionally, LLMs possess inherent knowledge of vulnerabilities from their training.
Therefore, we have designed three prompt schemes corresponding to three types of knowledge usage: \textit{LLM's own knowledge}, \textit{Raw knowledge}, and \textit{Summarized knowledge}.
\myfig~\ref{fig:scenarioprompt} illustrates how these types of knowledge can be combined with different CoT instructions to form three kinds of prompt schemes:

\begin{itemize}
\item In \textit{Scheme 1 - Raw}, we simply ask LLMs to generate results without any specific instructions. LLMs can use the APIs mentioned in \mysec\ref{sec:code} to retrieve related code segments. For open-source models, this scheme does not include this particular sentence.

\item In \textit{Scheme 2 - CoT}, we request LLMs to follow chain-of-thought instructions before generating the result. The LLMs should first summarize the functionality implemented by the given code segment, then analyze for any errors that could lead to vulnerabilities, and finally determine the vulnerability status.

\end{itemize}

\noindent
\textbf{Improved Instruction Following.}
Since all outputs from LLMs are in natural language, they are unstructured and need summarization and annotation to derive the final evaluation results. LLMs have been successfully utilized as evaluators~\cite{Chiang_Lee_2023,Li_Wang_Ma_Wu_Wang_Gao_Liu_2023}, and we use GPT-4.1 to automatically annotate the outputs of LLMs. The function calling API is employed to transform unstructured answers into structured results.
Specifically, \name generates structured results based on the answers provided by different prompt schemes and LLMs. These results include two parts: whether the LLM considers the code to be vulnerable, and the rationale for its vulnerability or lack thereof. The specific prompt used for this process can be found in Appendix~\ref{sec:summarizePrompt}. Following this step, we automatically annotate all the outputs from LLMs.

\subsection{LLM-based Result Annotation and Analysis}
\label{sec:annotation}

The components mentioned above are designed to enhance LLMs' vulnerability reasoning capabilities.
However, there is also a need for a component that enables automatic evaluation of LLMs' reasoning on vulnerabilities.
As such, we have designed \name to perform LLM-based result annotation, which is used to obtain the final evaluation results after individual LLMs generate their raw output.

\begin{figure}[t!]
    \centering
    \includegraphics[width=\linewidth]{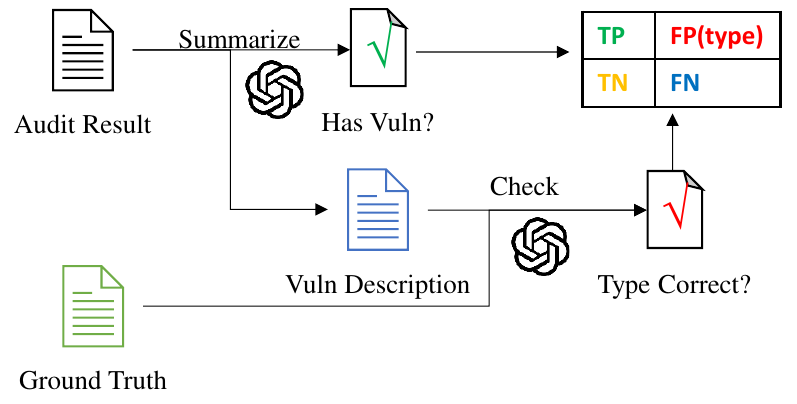}
    \vspace{-4ex}
    \caption{The process of automatic annotation by GPT-4.1.}
    \label{fig:annotation_process}
\end{figure}

Based on the raw Yes/No output from individual LLMs and whether the vulnerability type matches, as annotated by GPT-4.1, \name automatically obtains the following annotation results in terms of true positives, true negatives, false negatives, false positives, and false positive types:

\noindent \textbf{TP (True Positive):} The LLM correctly identifies the vulnerability with the correct type.

\noindent \textbf{TN (True Negative):} The LLM correctly concludes that the code is not vulnerable.

\noindent \textbf{FN (False Negative):} The LLM incorrectly identifies a vulnerable code segment as non-vulnerable.

\noindent \textbf{FP (False Positive):} The LLM incorrectly identifies a non-vulnerable code segment as vulnerable.

\noindent \textbf{FP\textsubscript{t} (False Positive Type):} The LLM identifies a vulnerable code segment as vulnerable, but with an incorrect vulnerability type.

Since FP-type includes both false positives (reporting a non-existent vulnerability) and false negatives (failing to report an existing vulnerability), we calculate the precision and recall of the LLMs' vulnerability detection results as follows:

\begin{equation}
    Precision = \frac{TP}{TP + FP + FP_{t}}
\end{equation}
\begin{equation}
    Recall = \frac{TP}{TP + FN + FP_{t}}
\end{equation}

For the accuracy of the LLMs' auto-annotated results, we randomly sampled 100 cases from each programming language and manually checked the results.
For the annotated results of binary classification (whether the code is vulnerable), we found that GPT-4.1 could reach 100\% accuracy across all three programming languages.
For the annotated results of whether the detected vulnerability type is same with ground truth, GPT-4.1 achieved an accuracy of 81\% for Solidity, 98\% for Java, and 97\% for C/C++.

%% file: implement.tex
\section{The \dataset Benchmark}
\label{sec:implement}

To enable \name to systematically evaluate LLMs' vulnerability reasoning capabilities under controlled conditions, a suitable benchmark dataset is necessary. 
While several recent datasets, such as PrimeVul~\cite{ding2024vulnerability} and MegaVul~\cite{ni2024megavul}, have been developed specifically for evaluating LLMs on vulnerability detection, they are not directly compatible with our framework.
These benchmarks typically lack two critical components: (1) explicit vulnerability knowledge that can be retrieved for in-context learning, and (2) surrounding code context that supplements LLM understanding.
To address these limitations, we introduce \dataset, the first benchmark designed with both \emph{knowledge-retrievability} and \emph{context-supplementability} in mind.
In the rest of this section, we detail the construction of \dataset. %

\noindent
\textbf{Targeted Programming Languages.}
In \dataset, we evaluate LLMs' vulnerability reasoning across both traditional programming languages, such as C/C++ and Java, and smart contract languages, namely Solidity.
This selection allows us to cover a broad spectrum of vulnerability types, from low-level memory and type errors to high-level logic bugs. 
For Java and C/C++, the most prevalent vulnerabilities are tied to language-level features, such as unsafe memory operations, insecure input handling, and unsafe deserialization.
In contrast, smart contracts, particularly those written in Solidity~\cite{Solidity}, are more susceptible to business logic vulnerabilities~\cite{zhang2023demystifying,Sun_Wu_Xue_Liu_Wang_Xu_Xie_Liu_2023}, which are significantly underrepresented in the pre-training corpus of most LLMs~\cite{difisok24} due to the emerging and domain-specific nature of decentralized applications.

\noindent
\textbf{Principle for Knowledge Synthesis.}
To assess how external knowledge enhances LLMs' reasoning, we design \dataset to include \emph{retrievable knowledge} derived from past vulnerabilities.
Knowledge must come from past vulnerabilities because we need to simulate realistic detection scenarios and thus cannot use the corresponding vulnerability knowledge of a give code segment to test itself. 
Accordingly, for each programming language, we construct two separate datasets: (i) a ``knowledge set'' for building a retrievable database of vulnerability knowledge, and (ii) a ``testing set'' for evaluating LLM performance under various configurations.
Each language adopts a knowledge source that best reflects real-world developer-facing resources.
That said, for Solidity, we can leverage various vulnerability auditing reports while for Java and C/C++, which lack structured audit reports, we rely primarily on the Common Weakness Enumeration (CWE) taxonomy and the Common Vulnerabilities and Exposures (CVE) database to synthesize structured knowledge.

\begin{table}[!t]
    \caption{Statistics of the \dataset benchmark.}
    \vspace{-2ex}
    \label{tab:dataset}
    \centering
    \resizebox{\linewidth}{!}{
    \begin{tabular}{lrrll}
        \toprule
        \textbf{Dataset} & \textbf{Samples} & \textbf{Projects} & \textbf{Period} \\
        \midrule
        Solidity Knowledge set   & 1,013   & 251 & Jan 2021 - Jul 2023 \\
        Solidity Testing set     & 51 + 51 & 11  & Aug 2023 - Jan 2024 \\
        Java Knowledge Set       & 770     & N/A & N/A \\
        Java Testing Set         & 46 + 46 & N/A & Jan 2013 - Dec 2022\\
        C/C++ Knowledge Set      & 860     & N/A & N/A \\
        C/C++ Testing Set        & 50 + 50 & N/A & Jan 2013 - Dec 2022\\
        \bottomrule
    \end{tabular}
    }
\end{table}

\noindent
\textbf{Detailed Collection and Synthesis.}
Table~\ref{tab:dataset} summarizes the datasets collected for Solidity, Java, and C/C++. 
For Solidity, we gathered high-risk vulnerabilities from the Code4rena (\cfr) bug bounty platform~\cite{noauthor_code4rena_2024}, specifically extracting 1,013 vulnerabilities from GitHub issues between January 2021 and July 2023~\cite{noauthor_C4GitHub_2024}.
These vulnerabilities and their associated reports comprise the knowledge set.
The testing set includes 51 vulnerable and 51 non-vulnerable code segments sourced from projects audited after July 2023. %
Non-vulnerable segments were randomly sampled from the same projects, with similar code length and complexity as the vulnerable counterparts.

For Java and C/C++, we relied on the CWE and CVE databases.
Specifically, we extracted 77 CWE categories for Java~\cite{CWEJAVA} and 86 for C/C++~\cite{CWECPP}.
Since CWEs offer high-level descriptions rather than code-aligned knowledge, we performed additional synthesis:
for each CWE, we prompted GPT-4.1 to generate 10 representative code examples, followed by generating corresponding audit-style vulnerability reports.
This resulted in 770 knowledge items for Java and 860 for C/C++.
To align the Java/C++ sets with the Solidity set, we curated 46 Java and 50 C/C++ vulnerabilities from the CVE and BigVul~\cite{BigVul} datasets, each paired with its patched or non-vulnerable version.

\noindent
\textbf{Knowledge Retrieval.}
After synthesizing the knowledge, we use FAISS~\cite{Johnson_Douze_Jegou_2017} to construct the vector database, and we set the top-K to retrieve the top-3 most relevant pieces of knowledge per query.
In FAISS, the query is embedded and calculates the dot product with all the vectors in the database; the top-K vectors with the highest dot products are returned as the result.
Moreover, the case without knowledge will be executed 3 times, which is the same as the case with knowledge.
As a result, for all three programming languages, although there are only 147 vulnerable cases and 147 non-vulnerable cases, they generate diverse test cases for different combinations of seven knowledge retrievals (top-3 are retrieved for each type), two prompt schemes, and two context variations, amounting to a total of 3,528 ($294 \times 3 \times 2 \times 2$) scenarios and 8,232 ($294 \times 7 \times 2 \times 2$) test cases to evaluate each model.

\noindent
\textbf{Minimizing Data Leakage.}
To ensure the integrity and fairness of evaluation, we take precautions to minimize the risk of data leakage from pre-training corpora.
All testing data is semantically preserved but systematically rewritten to prevent lexical overlap with public sources.
Specifically, we use GPT-4.1 to rename all function names, variable names, and comments while strictly maintaining the original program semantics.
Importantly, the underlying code statements remain unchanged to preserve vulnerability behavior.
For all three languages, there are cases that appeared earlier than the typical knowledge cut-off dates of current LLMs, such as QwQ-32B, which was trained on data available before November 2024.
To mitigate this risk, all testing samples undergo the aforementioned rewriting process to reduce memorization effects.
We also retain the evaluation results of the non-sanitized version for Solidity in Appendix~\ref{sec:detailedresults} for comparison, while the data shown in \mysec\ref{sec:evaluate} are all sanitized.
In this way, our benchmark sets a new standard for minimizing data leakage in LLM-based vulnerability detection, as compared with prior studies~\cite{Gao_Wang_Zhou_Zhu_Zhang_2023, Khare_Dutta_Li_Solko-Breslin_Alur_Naik_2023, Ullah_Han_Pujar_Pearce_Coskun_Stringhini_2023, ding2024vulnerability}.

%% file: evaluate.tex
\section{Evaluation}
\label{sec:evaluate}

\newcommand{\dataUpC}{$\red{\bigtriangleup}$}
\newcommand{\dataUpR}{$\red{\uparrow}$}
\newcommand{\dataDownC}{$\green{\bigtriangledown}$}
\newcommand{\dataDownR}{$\green{\downarrow}$}

\definecolor{colorTP}{RGB}{ 34, 139,  34}   %
\definecolor{colorTN}{RGB}{ 30, 144, 255}   %
\definecolor{colorFP}{RGB}{220,  20,  60}   %
\definecolor{colorFN}{RGB}{255, 165,   0}   %
\definecolor{colorFPt}{RGB}{148,   0, 211}   %

\newcommand{\FiveBar}[5]{%
  \pgfmathparse{(#1+#2+#3+#4+#5)}%
  \let\totalValue\pgfmathresult%
  \ifdim\totalValue pt=0pt
    \def\totalValue{1}%
  \fi
  \pgfmathsetlengthmacro{\lenTP}{(#1/\totalValue)*0.2\linewidth}%
  \pgfmathsetlengthmacro{\lenTN}{(#2/\totalValue)*0.2\linewidth}%
  \pgfmathsetlengthmacro{\lenFP}{(#3/\totalValue)*0.2\linewidth}%
  \pgfmathsetlengthmacro{\lenFN}{(#4/\totalValue)*0.2\linewidth}%
  \pgfmathsetlengthmacro{\lenFPt}{(#5/\totalValue)*0.2\linewidth}%
  {%
    \color{colorTP}\rule{\lenTP}{1ex}%
    \color{colorTN}\rule{\lenTN}{1ex}%
    \color{colorFP}\rule{\lenFP}{1ex}%
    \color{colorFN}\rule{\lenFN}{1ex}%
    \color{colorFPt}\rule{\lenFPt}{1ex}%
  }%
}

By applying the \name framework to the \dataset benchmark, we are able to decouple LLMs' inherent vulnerability reasoning capability from externally enhanceable factors and evaluate how this capability can be improved when combined with other enhancements, namely knowledge enhancement, context supplementation, and prompt schemes. 
With that said, we aim to provide \textit{automatic and standardized assessments for a fair evaluation of different LLMs' inherent vulnerability reasoning}, regardless of whether the models are strong or weak, rather than invent new LLM-based tools to achieve high detection metrics in vulnerability detection.
In this section, we follow \underline{this evaluation principle} to conduct a set of experiments and interpret their results.

\noindent
\textbf{LLMs Evaluated and Their Configurations.}
For LLMs, we benchmark the commonly used proprietary and open-source models available to us at the time of our evaluation in 2025. %
Specifically, from the LLMs listed in Table~\ref{tab:models}, for traditional foundation models, we select GPT-4.1 as the state-of-the-art proprietary model and choose Phi-3-mini-128k and Llama-3-8B as the open-source models.
Likewise, for deep reasoning models, we choose o4-mini as the state-of-the-art proprietary model and select QwQ-32B and DeepSeek-R1 as the open-source models.
All these LLMs are widely used in the AI community.
More background about these models was introduced in \mysec\ref{sec:background}.

To access these models, we use the OpenAI API~\cite{noauthor_function_2024} to interact with GPT-4.1 and o4-mini, and the Replicate API~\cite{replicate} to interact with all the open-source models.
We adhere to the default model configurations provided by the model providers,
except for the \texttt{temperature}, which we set to 0 for all models—except for o4-mini, which does not support temperature settings—as we aim to obtain the most deterministic results for reproducibility studies.
Moreover, the seeds are fixed and generated by a time-based randomness generator, which will be included in the artifact for replication.

\noindent
\textbf{Evaluation Metrics.}
For the metrics, we adhere to the standards specified in \mysec\ref{sec:annotation} for calculating True Positives (TP~{\color{colorTP}\rule{0.5em}{1ex}}), True Negatives (TN~{\color{colorTN}\rule{0.5em}{1ex}}), False Positives (FP~{\color{colorFP}\rule{0.5em}{1ex}}), False Negatives (FN~{\color{colorFN}\rule{0.5em}{1ex}}), and False Positive Type (FP\textsubscript{t}~{\color{colorFPt}\rule{0.5em}{1ex}}).
More specifically, larger TP and TN values (i.e., longer~{\color{colorTP}\rule{0.5em}{1ex}} and~{\color{colorTN}\rule{0.5em}{1ex}}) indicate better performance, while larger FP and FN values (i.e., longer~{\color{colorFP}\rule{0.5em}{1ex}} and~{\color{colorFN}\rule{0.5em}{1ex}}) suggest worse performance.
On the other hand, FP\textsubscript{t}~{\color{colorFPt}\rule{0.5em}{1ex}} is more borderline, indicating that the LLM detects a ground-truth vulnerability but misidentifies its root cause.

\noindent
\textbf{Overall Results.}
Tables~\ref{tab:bar_table_Solidity}, \ref{tab:bar_table_Java}, and \ref{tab:bar_table_C/C++} present the LLMs' vulnerability analysis results\footnote{The detailed raw results, including the calculated precision, recall, and F1 scores, are available in Table~\ref{tab:evaluation_all_detail_solidity} and Table~\ref{tab:evaluation_all_detail_java_cpp}; see Appendix~\ref{sec:detailedresults}.} for Solidity, Java, and C/C++, respectively, under different combinations of knowledge and context: ``Nk'' refers to results with no extra knowledge, ``Ok'' to results with original reports, ``Sk'' to results with summarized knowledge, ``C'' to results with context, and ``N'' to results without context.
It is worth noting that these results are obtained under the raw prompt scheme; the effect of CoT will be evaluated mainly in \mysec\ref{sec:RQ3}.
Based on these results, we perform a correlation analysis and address the corresponding research questions (RQs) in \mysec\ref{sec:RQ1} and \mysec\ref{sec:RQ2}.
Lastly, we conduct a pilot study in \mysec\ref{sec:RQ6} to demonstrate the practicality of \name's evaluation findings for zero-day vulnerability discovery in bug bounty projects.

\begin{table}[t]
\centering
\caption{\underline{Solidity} vulnerability analysis results of TP~{\color{colorTP}\rule{0.5em}{1ex}}, TN~{\color{colorTN}\rule{0.5em}{1ex}}, FP~{\color{colorFP}\rule{0.5em}{1ex}}, FN~{\color{colorFN}\rule{0.5em}{1ex}}, FP\textsubscript{t}~{\color{colorFPt}\rule{0.5em}{1ex}} under different combinations of knowledge (Nk/Ok/Sk) \& context (C/N) for different LLMs.}
    \vspace{-2ex}
\label{tab:bar_table_Solidity}
\resizebox{\linewidth}{!}{
\begin{tabular}{cllcll}
\toprule
\textbf{M} & \textbf{Setup} & \textbf{Metrics} & \textbf{M} & \textbf{Setup} & \textbf{Metrics} \\
\midrule

\multirow{6}{*}{\rotatebox{90}{GPT-4.1}} & NkC & \FiveBar{2}{32}{19}{28}{21} & \multirow{6}{*}{\rotatebox{90}{o4-mini}} & NkC & \FiveBar{14}{17}{34}{9}{28} \\
 & NkN & \FiveBar{3}{32}{19}{35}{13} & & NkN & \FiveBar{11}{18}{33}{15}{25} \\
 & OkC & \FiveBar{18}{106}{47}{94}{41} & & OkC & \FiveBar{20}{126}{27}{110}{23} \\
 & OkN & \FiveBar{16}{100}{53}{103}{34} & & OkN & \FiveBar{13}{132}{21}{123}{17} \\
 & SkC & \FiveBar{15}{134}{19}{123}{15} & & SkC & \FiveBar{20}{128}{25}{112}{21} \\
 & SkN & \FiveBar{13}{133}{20}{121}{19} & & SkN & \FiveBar{19}{126}{27}{110}{24} \\
\multirow{6}{*}{\rotatebox{90}{Llama-3-8B}} & NkC & \FiveBar{8}{0}{51}{0}{43} & \multirow{6}{*}{\rotatebox{90}{DeepSeek-R1}} & NkC & \FiveBar{13}{0}{51}{1}{37} \\
 & NkN & \FiveBar{6}{0}{51}{0}{45} & & NkN & \FiveBar{8}{5}{46}{1}{42} \\
 & OkC & \FiveBar{26}{1}{152}{2}{125} & & OkC & \FiveBar{34}{39}{114}{27}{92} \\
 & OkN & \FiveBar{27}{4}{149}{4}{122} & & OkN & \FiveBar{34}{48}{105}{35}{84} \\
 & SkC & \FiveBar{41}{7}{146}{1}{111} & & SkC & \FiveBar{36}{42}{111}{30}{87} \\
 & SkN & \FiveBar{38}{5}{148}{4}{111} & & SkN & \FiveBar{36}{49}{104}{31}{86} \\
\multirow{6}{*}{\rotatebox{90}{Phi-3-mini-128k}} & NkC & \FiveBar{10}{0}{51}{0}{41} & \multirow{6}{*}{\rotatebox{90}{QwQ-32B}} & NkC & \FiveBar{29}{4}{47}{2}{20} \\
 & NkN & \FiveBar{11}{0}{51}{0}{40} & & NkN & \FiveBar{34}{2}{49}{1}{16} \\
 & OkC & \FiveBar{44}{1}{152}{0}{109} & & OkC & \FiveBar{52}{52}{101}{61}{40} \\
 & OkN & \FiveBar{40}{1}{152}{0}{113} & & OkN & \FiveBar{41}{75}{78}{72}{40} \\
 & SkC & \FiveBar{40}{0}{153}{0}{113} & & SkC & \FiveBar{51}{73}{80}{59}{43} \\
 & SkN & \FiveBar{54}{0}{153}{0}{99} & & SkN & \FiveBar{47}{89}{64}{55}{51} \\

\bottomrule
\end{tabular}
}
\end{table}

\begin{table}[t]
\centering
\caption{\underline{Java} vulnerability analysis results of TP~{\color{colorTP}\rule{0.5em}{1ex}}, TN~{\color{colorTN}\rule{0.5em}{1ex}}, FP~{\color{colorFP}\rule{0.5em}{1ex}}, FN~{\color{colorFN}\rule{0.5em}{1ex}}, FP\textsubscript{t}~{\color{colorFPt}\rule{0.5em}{1ex}} under different combinations of knowledge (Nk/Ok/Sk) and context (C/N) for different LLMs.}
    \vspace{-2ex}
\label{tab:bar_table_Java}
\resizebox{\linewidth}{!}{
\begin{tabular}{cllcll}
\toprule
\textbf{M} & \textbf{Setup} & \textbf{Metrics} & \textbf{M} & \textbf{Setup} & \textbf{Metrics} \\
\midrule

\multirow{6}{*}{\rotatebox{90}{GPT-4.1}} & NkC & \FiveBar{15}{93}{45}{93}{30} & \multirow{6}{*}{\rotatebox{90}{o4-mini}} & NkC & \FiveBar{54}{33}{105}{27}{57} \\
 & NkN & \FiveBar{18}{93}{45}{93}{27} & & NkN & \FiveBar{42}{24}{114}{24}{72} \\
 & OkC & \FiveBar{2}{106}{32}{102}{34} & & OkC & \FiveBar{3}{112}{26}{115}{20} \\
 & OkN & \FiveBar{2}{107}{31}{103}{33} & & OkN & \FiveBar{1}{124}{14}{122}{15} \\
 & SkC & \FiveBar{3}{105}{33}{100}{35} & & SkC & \FiveBar{1}{95}{43}{99}{38} \\
 & SkN & \FiveBar{2}{108}{30}{102}{34} & & SkN & \FiveBar{6}{110}{28}{101}{31} \\
\multirow{6}{*}{\rotatebox{90}{Llama-3-8B}} & NkC & \FiveBar{39}{0}{138}{0}{99} & \multirow{6}{*}{\rotatebox{90}{DeepSeek-R1}} & NkC & \FiveBar{39}{0}{138}{0}{99} \\
 & NkN & \FiveBar{36}{0}{138}{0}{102} & & NkN & \FiveBar{42}{3}{135}{0}{96} \\
 & OkC & \FiveBar{4}{12}{126}{8}{126} & & OkC & \FiveBar{4}{93}{45}{91}{43} \\
 & OkN & \FiveBar{6}{10}{128}{7}{125} & & OkN & \FiveBar{3}{95}{43}{98}{37} \\
 & SkC & \FiveBar{13}{2}{136}{0}{125} & & SkC & \FiveBar{9}{64}{74}{60}{69} \\
 & SkN & \FiveBar{9}{1}{137}{0}{129} & & SkN & \FiveBar{8}{61}{77}{66}{64} \\
\multirow{6}{*}{\rotatebox{90}{Phi-3-mini-128k}} & NkC & \FiveBar{18}{0}{138}{0}{120} & \multirow{6}{*}{\rotatebox{90}{QwQ-32B}} & NkC & \FiveBar{63}{12}{126}{12}{63} \\
 & NkN & \FiveBar{21}{0}{138}{0}{117} & & NkN & \FiveBar{69}{12}{126}{12}{57} \\
 & OkC & \FiveBar{7}{4}{134}{4}{127} & & OkC & \FiveBar{7}{105}{33}{102}{29} \\
 & OkN & \FiveBar{5}{4}{134}{0}{133} & & OkN & \FiveBar{3}{111}{27}{105}{30} \\
 & SkC & \FiveBar{16}{1}{137}{2}{120} & & SkC & \FiveBar{16}{80}{58}{74}{48} \\
 & SkN & \FiveBar{12}{1}{137}{0}{126} & & SkN & \FiveBar{7}{78}{60}{83}{48} \\

\bottomrule
\end{tabular}
}
\end{table}

\begin{table}[t]
\centering
\caption{\underline{C/C++} vulnerability analysis results of TP~{\color{colorTP}\rule{0.5em}{1ex}}, TN~{\color{colorTN}\rule{0.5em}{1ex}}, FP~{\color{colorFP}\rule{0.5em}{1ex}}, FN~{\color{colorFN}\rule{0.5em}{1ex}}, FP\textsubscript{t}~{\color{colorFPt}\rule{0.5em}{1ex}} under different combinations of knowledge (Nk/Ok/Sk) \& context (C/N) for different LLMs.}
    \vspace{-2ex}
\label{tab:bar_table_C/C++}
\resizebox{\linewidth}{!}{
\begin{tabular}{cllcll}
\toprule
\textbf{M} & \textbf{Setup} & \textbf{Metrics} & \textbf{M} & \textbf{Setup} & \textbf{Metrics} \\
\midrule

\multirow{6}{*}{\rotatebox{90}{GPT-4.1}} & NkC & \FiveBar{45}{54}{96}{54}{51} & \multirow{6}{*}{\rotatebox{90}{o4-mini}} & NkC & \FiveBar{39}{33}{117}{24}{87} \\
 & NkN & \FiveBar{36}{60}{90}{57}{57} & & NkN & \FiveBar{54}{36}{114}{33}{63} \\
 & OkC & \FiveBar{22}{96}{54}{94}{34} & & OkC & \FiveBar{21}{123}{27}{113}{16} \\
 & OkN & \FiveBar{12}{98}{52}{91}{47} & & OkN & \FiveBar{16}{124}{26}{116}{18} \\
 & SkC & \FiveBar{16}{99}{51}{90}{44} & & SkC & \FiveBar{24}{109}{41}{93}{33} \\
 & SkN & \FiveBar{10}{93}{57}{92}{48} & & SkN & \FiveBar{18}{103}{47}{101}{31} \\
\multirow{6}{*}{\rotatebox{90}{Llama-3-8B}} & NkC & \FiveBar{30}{0}{150}{0}{120} & \multirow{6}{*}{\rotatebox{90}{DeepSeek-R1}} & NkC & \FiveBar{63}{0}{150}{0}{87} \\
 & NkN & \FiveBar{27}{0}{150}{0}{123} & & NkN & \FiveBar{45}{0}{150}{0}{105} \\
 & OkC & \FiveBar{13}{0}{150}{0}{137} & & OkC & \FiveBar{20}{79}{71}{73}{57} \\
 & OkN & \FiveBar{5}{0}{150}{0}{145} & & OkN & \FiveBar{16}{69}{81}{66}{68} \\
 & SkC & \FiveBar{34}{1}{149}{1}{115} & & SkC & \FiveBar{33}{53}{97}{43}{74} \\
 & SkN & \FiveBar{32}{0}{150}{1}{117} & & SkN & \FiveBar{31}{41}{109}{43}{76} \\
\multirow{6}{*}{\rotatebox{90}{Phi-3-mini-128k}} & NkC & \FiveBar{39}{0}{150}{0}{111} & \multirow{6}{*}{\rotatebox{90}{QwQ-32B}} & NkC & \FiveBar{99}{21}{129}{9}{42} \\
 & NkN & \FiveBar{36}{0}{150}{0}{114} & & NkN & \FiveBar{72}{27}{123}{9}{69} \\
 & OkC & \FiveBar{19}{2}{148}{0}{131} & & OkC & \FiveBar{27}{83}{67}{81}{42} \\
 & OkN & \FiveBar{19}{0}{150}{0}{131} & & OkN & \FiveBar{24}{81}{69}{80}{46} \\
 & SkC & \FiveBar{26}{4}{146}{3}{121} & & SkC & \FiveBar{43}{62}{88}{52}{55} \\
 & SkN & \FiveBar{37}{3}{147}{2}{111} & & SkN & \FiveBar{35}{61}{89}{57}{58} \\

\bottomrule
\end{tabular}
}
\end{table}

\input{RQ1}

\input{RQ2}

\input{RQ3}

\input{RQ6}

%% file: RQ1.tex
\subsection{RQ1: Effects of Knowledge Enhancement}
\label{sec:RQ1}

In this RQ, we evaluate LLMs' inherent vulnerability reasoning with and without the effects of the knowledge enhancement mechanisms introduced in \mysec\ref{sec:knowledge}.

\noindent
\textbf{Results for Solidity.}
According to the high-level Table~\ref{tab:bar_table_Solidity} and the detailed Table~\ref{tab:evaluation_all_detail_solidity} (see Appendix~\ref{sec:detailedresults}), both types of knowledge (i.e., original reports and summarized knowledge) supplements improve vulnerability reasoning for Solidity when applied to traditional foundation models such as GPT-4.1, Phi-3, and Llama-3.
On these models, both precision and recall are improved for most combinations.
For GPT-4.1 with CoT, the precision does not change significantly; however, the recall drops—from 33.33\% to 19.61\% and 11.76\%, and from 17.65\% to 6.54\% and 7.84\% for original reports and summarized knowledge, respectively.
In contrast, for deep reasoning models like DeepSeek-R1, there is no significant difference, and even a drop in precision and recall is observed in QwQ-32B.
For o4-mini, the precision increases by more than 10\% under combinations with the raw prompt scheme, while the precision does not change significantly under the CoT scheme.
As with other deep reasoning models, the recall for o4-mini decreases when external knowledge is provided.

When examining the number of true and false cases, we observe that for traditional foundation models using the raw scheme, the number of TP cases increases when any type of knowledge is provided, with all 12 out of 12 combinations showing an increase.
This situation changes under the CoT scheme: only 6 out of 12 combinations show an increase in TP, while the remaining 6 combinations show a decrease.
Another significant change is observed in the number of TN and FN cases.
Since Phi-3 always classifies all code as vulnerable, the numbers of TN and FN cases are nearly 0 and will not be discussed further.
For all other models, including both traditional foundation and deep reasoning models, both TNs and FNs increase when knowledge is provided.
On DeepSeek-R1, the increase is dramatic—rising from nearly 0 to around 40.
For o4-mini, the numbers of TN and FN cases nearly double when summarized knowledge is provided.
Compared to original reports, summarized knowledge has a more significant impact on the number of TN cases, with 18 out of 20 combinations increasing TN, while 13 out of 20 combinations show an increase in FN for original reports.

\noindent
\textbf{Results for Java and C/C++.}
However, for Java and C/C++, some of the above findings do not hold.
According to the high-level Tables~\ref{tab:bar_table_Java} and \ref{tab:bar_table_C/C++} and the detailed Table~\ref{tab:evaluation_all_detail_java_cpp} (see Appendix~\ref{sec:detailedresults}), with respect to the number of TP cases and recalls, we observe that for these two languages the trend is similar to that observed in the deep reasoning models on Solidity.
In all combinations, the number of TP cases decreases when any type of knowledge is provided, with 89 out of 96 combinations showing a decrease (and one remaining unchanged).
For the TN and FN cases, the trend is similar: the provision of knowledge increases both TN and FN cases.
However, compared to summarized knowledge, which results in the most significant increases in TN and FN, original reports yield the highest TN in 87 out of 96 combinations and the highest FN in 91 out of 96 combinations.
Consequently, for precision, which is affected by decreases in both TP and FP cases, most combinations (80 out of 96) show a decline when external knowledge is supplied compared to when it is not.

\noindent
\textbf{Comparison of both results.}
Based on the above findings, we conclude that knowledge enhancement does not affect different programming languages or model types consistently.
For deep reasoning models, the provision of external knowledge does not consistently improve performance across all three languages and may even harm it.
Our manual examination of the reasoning process in deep reasoning models reveals that they use more tokens to confirm the existence of all potential vulnerability types.
When incorrect knowledge is provided, the model may overlook other potential types by focusing on the erroneous one, thereby producing a negative result.

In contrast, for traditional foundation models, summarized knowledge significantly improves vulnerability reasoning for Solidity.
This is because vulnerabilities in Solidity are more logic-based and require functionality-based knowledge retrieval to provide the relevant information to the LLMs.
For Java and C/C++, most vulnerabilities are related to intrinsic language features, such as unsafe deserialization and command injection; with this advantage, traditional foundation models that are supplied with proper external knowledge can achieve better performance on Solidity.

Another notable finding is that all types of knowledge provision increase the number of negative cases, which is counter-intuitive.
In the early stages of LLM-based vulnerability detection~\cite{David_Zhou_Qin_Song_Cavallaro_Gervais_2023}, large language models tended to assume that vulnerabilities mentioned in the prompts would exist in the code.
This finding suggests that current LLMs are more inclined to reason about the conditions necessary for a vulnerability to exist rather than focusing on the specific vulnerability descriptions in the prompts.

\finding{1}{
Three sub-findings regarding the effects of knowledge enhancement on LLMs' inherent vulnerability reasoning:\\
\textbf{(a)} For traditional foundation models, when dealing with logic-based vulnerabilities (e.g., Solidity), summarized knowledge significantly improves vulnerability reasoning, whereas for traditional languages (e.g., Java and C/C++), it is preferable to rely on the LLMs' built-in knowledge;\\
\textbf{(b)} Knowledge enhancement on deep reasoning models, on the other hand, has similar impacts across different programming languages, surprisingly often harming precision, recall, and F1-score;\\
\textbf{(c)} Across all languages, the provision of external knowledge increases the number of negative cases, indicating that current LLMs are more inclined to reason about the conditions necessary for a vulnerability to exist rather than focus on the specific vulnerability descriptions in the prompts.
}

%% file: RQ2.tex
\subsection{RQ2: Effects of Context Supplementation}
\label{sec:RQ2}

In this RQ, we evaluate LLMs' inherent vulnerability reasoning with and without the effects of context supplementation, as introduced in \mysec\ref{sec:code}.
Tables~\ref{tab:bar_table_Solidity}, \ref{tab:bar_table_Java}, and \ref{tab:bar_table_C/C++} summarize the high-level results for all model–knowledge combinations, with context supplementation (``setup'' ending in ``C'') and without (``setup'' ending in ``N'').
Detailed results can be found in Tables~\ref{tab:evaluation_all_detail_solidity} and \ref{tab:evaluation_all_detail_java_cpp} in Appendix~\ref{sec:detailedresults}.

Across all three programming languages, context supplementation shows a slight and generally positive effect on the number of TPs and recall, though the impact is not consistent.
For Solidity, 22 out of 36 model–prompt–knowledge combinations show an increase in TPs with context, while 11 show a decrease.
In Java, 18 combinations show an increase and 15 a decrease.
C/C++ sees the most consistent improvement, with 28 combinations showing increased TPs and only 7 showing a decrease.
These shifts in TPs are mirrored in the recall trends.
As for TNs, FPs, and FNs, the changes vary without a clear pattern that context supplementation does not consistently improve or degrade these metrics across languages or model types.

The precision metric follows a similar trend.
In Solidity, precision improves in 22 combinations and drops in 13.
In Java, 19 combinations improve while 17 drop.
In C/C++, 29 out of 36 combinations show improved precision, while only 7 show a decrease.
This suggests that context supplementation tends to slightly enhance precision, though not uniformly.

An interesting observation is that for traditional foundation models, the highest F1 scores often occur when context is supplemented.
In Solidity, 10 out of 18 combinations yield higher F1 scores with context, while 7 combinations see a decrease.
For Java and C/C++, similar proportions are observed, although in C/C++ the highest absolute F1 score is achieved by QwQ-32B without context supplementation.
This implies that while context supplementation generally helps foundation models better understand code semantics, deep reasoning models may already extract sufficient information from the isolated code segment and thus benefit less—or even be slightly hindered—by the addition of surrounding context.

\finding{2}{
Context supplementation offers slight and inconsistent improvements in LLMs' vulnerability reasoning across precision, recall, and F1 scores. While traditional foundation models often achieve their highest F1 scores with context, deep reasoning models tend to perform best without it, indicating that their reasoning may already leverage sufficient internal code semantics without additional context.
}

%% file: RQ3.tex
\subsection{RQ3: Effects of Different Prompt Schemes}
\label{sec:RQ3}

In this RQ, we evaluate LLMs' inherent vulnerability reasoning under different prompt schemes, focusing specifically on Chain-of-Thought (CoT) prompts as introduced in \mysec\ref{sec:scheme}.
According to the detailed results shown in Tables~\ref{tab:evaluation_all_detail_solidity} and \ref{tab:evaluation_all_detail_java_cpp} (see Appendix~\ref{sec:detailedresults}), CoT prompts can sometimes improve performance over raw prompts, but the effects are not consistent across all models and languages.

For TP cases, CoT prompts lead to improvements in only 11 out of 18 combinations for traditional foundation models and 7 out of 18 for deep reasoning models on Solidity.
A similar pattern is observed on Java and C/C++, where CoT improves TPs in 7 and 11 out of 18 combinations for general-purpose models, and 9 out of 18 for deep reasoning models in both languages.
These results indicate that CoT does not reliably increase TP counts.

In contrast, the impact of CoT prompts on FPs and TNs is more favorable.
On Solidity and Java, 12 out of 18 combinations show reduced FPs and increased TNs for both traditional foundation and deep reasoning models.
For C/C++, 10 out of 18 combinations show improvement in FPs and TNs for traditional foundation models, and 11 out of 18 for deep reasoning models.
This suggests that CoT prompts are more effective at reducing incorrect positive predictions and enhancing correct negative predictions.

A notable exception is Phi-3, where the CoT prompt scheme causes degradation across all metrics—FPs, TNs, and FNs—across all programming languages. This is because Phi-3 tends to classify nearly all samples as vulnerable, resulting in negligible FN but high FP, and CoT fails to correct this behavior.

Overall, while CoT prompts are not a guaranteed solution for improving recall or TP count, they tend to offer moderate gains in reducing false positives and increasing true negatives, particularly for deep reasoning models.

\finding{3}{
While CoT prompt schemes do not consistently improve recall or the number of true positives,
they tend to reduce false positives and increase true negatives across most models and languages, making them a beneficial prompt strategy for improving prediction reliability in vulnerability detection.
}

%% file: RQ6.tex
\subsection{RQ4: Testing for Zero-Day Vulnerabilities}
\label{sec:RQ6}

To demonstrate the usefulness of \name's evaluation findings in identifying new vulnerabilities, we have deployed a Solidity-specific version of \name to our industry partner, a Web3 security company.
This is because knowledge enhancement shows the most significant benefits for Solidity due to its prevalence of logic-based vulnerabilities, as revealed in RQ1.
In contrast, vulnerabilities in C/C++ and Java are more often tied to language-level features and are thus better handled by the LLMs' built-in knowledge.
Additionally, Solidity contracts tend to be smaller and more self-contained, making them suitable for direct LLM evaluation.
In comparison, C/C++/Java projects are often much larger in scale, rendering it infeasible to apply LLMs directly to all functions without first narrowing the scope.
We believe that combining LLMs with traditional techniques such as program analysis or fuzzing would be more appropriate for large-scale applications in those languages.
As a result, in this RQ, we focus on the zero-day vulnerabilities discovered by \name in Solidity projects only.

Specifically, we used \name for Solidity to audit four bug bounty smart contract projects—Apebond~\cite{Apebond}, Glyph AMM~\cite{Glyph}, StakeStone~\cite{Stakestone}, and Hajime—on~\cite{Hajime}.
After a brief manual review to filter out factual errors, all outputs were submitted to the Secure3 community for confirmation.
In total, we submitted 29 issues across these four projects, and 14 were confirmed by the community.
Details of the projects and bounties can be found in Table~\ref{tab:audit_projects}.
From these combined audits, we received a total bounty of \$3,576, demonstrating the practicality of \name.

In the rest of this section, we conduct case studies on the confirmed issues in the \textit{Apebond} project~\cite{SoulSolidity/SoulZapV1_2023}, which involves a set of contracts for single transaction token swaps, liquidity provision, and bond purchases.
The audit report for \textit{Apebond} is available at~\cite{ApebondReport} and does not include the real names of the auditors.
The retrieved knowledge can be found in our open-source repository~\cite{FourKnowledge}.

The first case involves an iteration without proper checks for duplicate entries, potentially leading to financial losses.
The source code is shown in \myfig~\ref{code:finding_1}.
The function \texttt{\_routerSwapFromPath} is designed to execute a token swap operation using the input parameter \texttt{\_uniSwapPath}, which is expected to contain a swapping path \texttt{\_uniSwapPath.path} indicating the series of token conversions to be executed.
If the input array contains duplicate entries, it will result in unnecessary token conversions, and the fees paid for these duplicated conversions will be lost.
Using \name, we matched the functionality of the code with the knowledge that ``For any functionality that involves processing input arrays, especially in smart contracts or systems managing assets and tokens, it's crucial to implement stringent validation mechanisms to check for duplicate entries.''
The full details of this knowledge are available in Knowledge 1~\cite{FourKnowledge}.

\begin{table}[t]
    \centering
    \caption{Four bounty Solidity projects audited in RQ4.}
    \vspace{-2ex}
    \label{tab:audit_projects}
    \begin{tabular}{lrrr}
        \toprule
        \textbf{Project} & \textbf{Bounty (\$)} & \textbf{Submitted} & \textbf{Confirmed} \\
        \midrule
        Apebond & 376 & 12 & 4 \\
        Glyph AMM & 329 & 6 & 4 \\
        StakeStone & 2,281 & 9 & 5 \\
        Hajime & 590 & 2 & 1 \\
        \midrule
        \textbf{Total} & 3,576 & 29 & 14 \\
        \bottomrule
    \end{tabular}
\end{table}

\begin{figure}[t]
\begin{lstlisting}[
        language=Solidity,
        linewidth=.48\textwidth,
        frame=none,
        xleftmargin=.03\textwidth,
        ]
function _routerSwapFromPath(
    SwapPath memory _uniSwapPath,
    uint256 _amountIn,
    address _to,
    uint256 _deadline
) private returns (uint256 amountOut) {
    require(_uniSwapPath.path.length >= 2, "SoulZap: need path0 of >=2");
    address outputToken = _uniSwapPath.path[_uniSwapPath.path.length - 1];
    uint256 balanceBefore = _getBalance(IERC20(outputToken), _to);
    _routerSwap(
        _uniSwapPath.swapRouter,
        _uniSwapPath.swapType,
        _amountIn,
        _uniSwapPath.amountOutMin,
        _uniSwapPath.path,
        _to,
        _deadline
    );
    amountOut = _getBalance(IERC20(outputToken), _to) - balanceBefore;
}
\end{lstlisting}
    \vspace{-3ex}
    \caption{Case 1: Lack of Duplication Check for Input Array.}
    \label{code:finding_1}
\end{figure}

The second case, as shown in \myfig~\ref{code:finding_2}, involves a precision calculation error that could lead to financial loss.
In lines 7 and 8, the function \texttt{getSwapRatio} calculates the underlying balances of the swap tokens.
This process includes normalization of precision, but it incorrectly assumes that the precision of the input token is always 18 decimal places.
However, when obtaining price ratios from other oracles, the precision may vary and not always be 18 decimal places.
This incorrect assumption about precision can lead to miscalculation of the swap ratio, potentially causing the user to gain tokens that do not belong to them or lose a number of tokens.
With \name, we matched the functionality of the code with the knowledge that ``The vulnerability stems from an incorrect handling of decimal precision while calculating the price ratio between two oracles with different decimals.''
The full details of this knowledge are available in Knowledge~2~\cite{FourKnowledge}.

\begin{figure}[t!]
\begin{lstlisting}[
        language=Solidity,
        linewidth=.48\textwidth,
        frame=none,
        xleftmargin=.03\textwidth,
        ]
function getSwapRatio(
    SwapRatioParams memory swapRatioParams
) internal view returns (uint256 amount0, uint256 amount1) {
    // More code
    vars.token0decimals = ERC20(address(swapRatioParams.token0)).decimals();
    vars.token1decimals = ERC20(address(swapRatioParams.token1)).decimals();
    vars.underlying0 = _normalizeTokenDecimals(vars.underlying0, vars.token0decimals);
    vars.underlying1 = _normalizeTokenDecimals(vars.underlying1, vars.token1decimals);
    // More code
}
function _normalizeTokenDecimals(
    uint256 amount,
    uint256 decimals
) internal pure returns (uint256) {
    return amount * 10 ** (18 - decimals);
}
\end{lstlisting}
    \vspace{-3ex}
    \caption{Case 2: Precision Calculation Error Type I.}
    \label{code:finding_2}
\end{figure}

\begin{figure}[t!]
\begin{lstlisting}[
        language=Solidity,
        linewidth=.48\textwidth,
        frame=none,
        xleftmargin=.03\textwidth,
        ]
function pairTokensAndValue(
    address token0,
    address token1,
    uint24 fee,
    address uniV3Factory
) internal view returns (uint256 price) {
    // More Code
    uint256 sqrtPriceX96;
    (sqrtPriceX96, , , , , , ) = IUniswapV3Pool(tokenPegPair).slot0();
    uint256 token0Decimals = getTokenDecimals(token0);
    uint256 token1Decimals = getTokenDecimals(token1);
    if (token1 < token0) price = (2 ** 192) / ((sqrtPriceX96) ** 2 / uint256(10 ** (token0Decimals + 18 - token1Decimals)));
    else price = ((sqrtPriceX96) ** 2) / ((2 ** 192) / uint256(10 ** (token0Decimals + 18 - token1Decimals)));
}
\end{lstlisting}
    \vspace{-3ex}
    \caption{Case 3: Precision Calculation Error Type II.}
    \label{code:finding_3}
\end{figure}

Similarly, as shown in \myfig~\ref{code:finding_3}, the function \texttt{pairTokensAndValue} is responsible for calculating the price of tokens using \texttt{sqrtPriceX96} obtained from a UniswapV3Pool.
However, this function also erroneously assumes that the precision of \texttt{sqrtPriceX96} is always 18 decimal places, which could result in unplanned benefits or loss of funds.
The knowledge matched in \name for this scenario is ``The vulnerability occurs when calculating the squared root price of a position in a liquidity pool with tokens having different decimal values.''
While two matched knowledge from cases 2 and 3 are not identical - the latter specifically mentions the ``squared root price'' - they are semantically similar and both applicable to describing the same type of vulnerability.
The full details of this knowledge are available in Knowledge 3~\cite{FourKnowledge}.

\begin{figure}[t!]
\begin{lstlisting}[
        language=Solidity,
        linewidth=.48\textwidth,
        frame=none,
        xleftmargin=.03\textwidth,
        ]
function _zap(
    ZapParams memory zapParams,
    SwapPath memory feeSwapPath,
    bool takeFee
) internal whenNotPaused {
    // Verify and setup
    if (zapParams.liquidityPath.lpType == LPType.V2) {
        // some checks
        vars.amount0In = zapParams.amountIn / 2;
        vars.amount1In = zapParams.amountIn / 2;
    }
    // More code ......
    if (zapParams.liquidityPath.lpType == LPType.V2) {
        (vars.amount0Lp, vars.amount1Lp, ) = IUniswapV2Router02(
            zapParams.liquidityPath.lpRouter
        ).addLiquidity(
                zapParams.token0,
                zapParams.token1,
                // ......
            );
    }
    // More code
}
\end{lstlisting}
    \vspace{-3ex}
    \caption{Case 4: Funding Allocation Error.}
    \label{code:finding_4}
\end{figure}

In \myfig~\ref{code:finding_4}, the function \texttt{\_zap} divides the input amount (\texttt{amountIn}) equally between \texttt{amount0In} and \texttt{amount1In} on lines 9 and 10.
However, the actual token reserve ratio in the pool may not be 1:1.
This discrepancy can lead to an imbalanced provision of liquidity when \texttt{addLiquidity} is called on line 16, as the token pair might require a different ratio for optimal liquidity provision.
This vulnerability can disrupt the equilibrium of liquidity pools and cause traders to lose tokens.
The knowledge matched in \name for this case is ``The fundamental vulnerability occurs when liquidity providers add liquidity to a pool of two tokens, and the token amounts provided have different proportions as compared to the existing liquidity pool. The contract uses the smaller of these proportions to calculate the amount of LP tokens minted.''
While this knowledge does not precisely describe the vulnerability, as there is no smaller proportion in this specific case, it can still be useful in detecting the vulnerability when combined with the reasoning ability of LLMs.
The full details of this knowledge are available in Knowledge 4~\cite{FourKnowledge}.

From the four cases above, it is evident that knowledge supplements in \name for Solidity can help detect vulnerabilities, even if they do not exactly match the knowledge stored in the vector database.
Among these four cases, only the first can be detected by static analysis tools, while the remaining three are logic bugs closely tied to business logic.
With this positive result, we demonstrate that \name's evaluation findings can be leveraged to identify real-world project bugs that existing tools do not detect.

%% file: discussion.tex
\section{Discussion and Future Work}
\label{sec:discussion}

\input{lessons}

\input{threat}

%% file: lessons.tex
\subsection{Lessons Learned}
\label{sec:lessons}

In this section, we reflect on the key insights from our empirical study in \mysec\ref{sec:evaluate} and outline possible directions for improving LLM-assisted vulnerability reasoning.

\noindent\textbf{Knowledge Enhancement.}
Our study shows that knowledge enhancement plays an important but uneven role across different languages and model types.
Specifically, it significantly improves vulnerability reasoning for traditional foundation models when dealing with logic-based languages like Solidity, while it may degrade performance on C/C++ and Java.
This calls for more accurate knowledge retrieval mechanisms, as mismatched knowledge can mislead model reasoning.
In future work, more sophisticated semantic-preserving retrieval methods, e.g., graph-based representations or symbolic feature extraction, should be explored to enhance relevance and minimize noise.
However, on reasoning models, there is a consistent degradation in performance when external knowledge is introduced, indicating that these models may already possess sufficient built-in knowledge for vulnerability detection.

\noindent\textbf{Context Supplement.}
Supplying surrounding code context can slightly improve performance but introduces variability across tasks and models.
Irrelevant or excessive context may distract models or trigger false positives.
Thus, context supplementation should be applied carefully, preferably with filtering strategies or model-specific context windows, to balance informativeness and conciseness.

\noindent\textbf{Prompt Scheme Selection.}
CoT prompts show moderate gains, especially in reducing false positives and improving true negatives.
However, the improvements are not uniform.
Designing effective CoT prompts requires decomposing tasks into interpretable reasoning steps without overwhelming the model.
CoT also may orchestrate with other enhancement like knowledge supplement that a proper combination of CoT and other techniques can further improve the performance.
Future work should investigate automatic prompt tuning or retrieval-augmented CoT pipelines to further leverage this strategy.

\noindent\textbf{Model Selection.}
Instead of the conventional distinction between open-source and proprietary models, our study distinguishes between \emph{traditional foundation models} and \emph{deep reasoning models}.
While reasoning models exhibit strong built-in capabilities, their performance may suffer when noisy external knowledge is introduced.
In contrast, foundation models benefit more from carefully curated knowledge.
Therefore, the choice of LLM should depend on both the language being audited and the model's sensitivity to external enhancements.
Baseline vulnerability reasoning capability should be prioritized over model popularity.

%% file: threat.tex
\subsection{More Accurate Knowledge Retrieval}
\label{sec:threat}

To ensure fairness, our \dataset benchmark uses fixed knowledge supplies across cases.
We manually verified the accuracy of retrieval results and identify room for improvement in real-world usage.
For Solidity, 68 out of 100 sampled positive cases had knowledge aligned with the ground truth, indicating reasonable retrieval quality and partially explaining the improved precision and recall.
However, vulnerabilities requiring low-level reasoning (e.g., integer overflow) are underrepresented.
For Java, only 36 of 100 samples retrieved relevant knowledge, largely limited to XSS and SQL injection, while issues like SSRF and CSRF were rarely matched due to weaker functional correspondence.
This reflects the difficulty of capturing procedural or contextual vulnerability types in static descriptions.

Even though exact matching is difficult, we show in RQ4 that partial similarity can still positively influence model predictions.
Thus, future work should explore retrieval strategies that focus on behavior- or functionality-level similarity rather than surface pattern matching.

\subsection{Covering More Programming Languages}

Our current \dataset benchmark targets three representative languages: Solidity (logic-heavy smart contracts), Java, and C/C++ (general-purpose procedural languages).
However, the \name framework is designed to be language-agnostic in principle.

To extend it to other languages (e.g., Python, Rust, JavaScript), researchers can build tailored vulnerability knowledge bases and implement supporting tools such as AST parsing, call graph construction, or symbolic execution.
Given language-specific semantics, the tool-invoking modules must be adapted accordingly.
Our modular framework design allows such extensions and supports future studies on broader language ecosystems.

%% file: related.tex
\section{Related Work}
\label{sec:related}

\noindent\textbf{LLM-based Vulnerability Detection.}
Vulnerability detection has long been a core challenge in software security.
Traditional approaches typically rely on static rules or fuzzing-based testing, which often struggle to detect novel or complex vulnerabilities.
Recent advances in code-oriented large language models~\cite{Roziere_Gehring_Gloeckle_Sootla_Gat_Tan_Adi_Liu_Remez_Rapin_et, Thapa_Jang_Ahmed_Camtepe_Pieprzyk_Nepal_2022} have led to new research exploring LLMs for vulnerability detection.

Several works have proposed LLM-based vulnerability detectors. 
Thapa \etal~\cite{Thapa_Jang_Ahmed_Camtepe_Pieprzyk_Nepal_2022} evaluated the effectiveness of LLMs on standard vulnerability datasets.
Alqarni \etal~\cite{Alqarni} fine-tuned LLMs for vulnerability classification.
Tang \etal~\cite{Tang} combined graph representations and LLMs for function-level detection.
Hu \etal~\cite{Hu_Huang_Ilhan_Tekin_Liu_2023} used LLM role-playing for analyzing smart contracts.

Other researchers leverage LLMs to enhance fuzz testing. 
Deng \etal~\cite{Deng_Xia_Peng_Yang_Zhang_2023} proposed TitanFuzz for fuzzing DL libraries.
FuzzGPT~\cite{Deng_Xia_Yang_Zhang_Yang_Zhang_2023} synthesizes corner-case programs using LLMs.
Meng \etal~\cite{Meng_Mirchev_Bohme_Roychoudhury} developed ChatAFL for fuzzing network protocols.
Fuzz4All~\cite{Xia_Paltenghi_Tian_Pradel_Zhang_2024} generalizes this to broader domains.

There is also work combining LLMs with static analysis.
Sun \etal~\cite{Sun_Wu_Xue_Liu_Wang_Xu_Xie_Liu_2023} combined GPT with symbolic execution for smart contracts.
Li \etal~\cite{Li_Hao_Zhai_Qian_2023} integrated LLMs with traditional analysis pipelines.
However, these studies primarily focus on detecting vulnerabilities rather than assessing the reasoning capability of LLMs in making security-relevant judgments.
In contrast, \name provides a systematic framework for benchmarking and analyzing the reasoning behind LLM-based vulnerability assessments.

\noindent\textbf{Benchmarking LLMs' Vulnerability Detection Capabilities.}
Another line of work evaluates LLMs on benchmark datasets.
For example, Chen \etal~\cite{Chen_Su_Chen_Wang_Bi_Wang_Lin_Chen_Zheng_2023} tested LLMs on Solidity vulnerability detection tasks.
David \etal~\cite{David_Zhou_Qin_Song_Cavallaro_Gervais_2023} evaluated LLMs on real-world DeFi smart contracts.
Khare \etal~\cite{Khare_Dutta_Li_Solko-Breslin_Alur_Naik_2023} benchmarked LLMs on Java and C++.
Gao \etal~\cite{Gao_Wang_Zhou_Zhu_Zhang_2023}, Ullah \etal~\cite{Ullah_Han_Pujar_Pearce_Coskun_Stringhini_2023}, and Ding \etal~\cite{ding2024vulnerability} constructed datasets for evaluating different detection methods across programming languages.
Lin \etal~\cite{Lin2023from} evaluated LLMs with different configurations, including quantization and context length, on vulnerability detection tasks.
Meanwhile, others examined LLMs in related tasks such as vulnerability repair~\cite{Pearce_Tan_Ahmad_Karri_Dolan-Gavitt_2023, zhang2024acfix}.
Unlike these efforts, which focus on raw performance metrics, our work decouples LLMs' reasoning capability from external enhancements like knowledge, context, and prompt design.
\dataset is the first benchmark designed to support such modular evaluations with retrievable knowledge and context-supplementable code.

\noindent\textbf{Security-oriented LLMs.}
There is a growing body of work on developing security-specific LLMs.
Lacomis \etal~\cite{Lacomis_Yin_Schwartz_Allamanis_Le_Goues_Neubig_Vasilescu_2019} and Pal \etal~\cite{len_index_count} built models for recovering variable names in binary code.
Pei \etal~\cite{Pei_Li_Jin_Liu_Geng_Cavallaro_Yang_Jana_2023} proposed transformers tailored for code semantics in security tasks.
Chen \etal~\cite{Chen_Lacomis_Schwartz_Goues_Neubig_Vasilescu_2022} improved decompilation for downstream analysis.
Ding \etal~\cite{Ding_Steenhoek_Pei_Kaiser_Le_Ray_2023} introduced execution-aware pretraining.
Gai \etal~\cite{Gai_Zhou_Qin_Song_Gervais_2023} and Guthula \etal~\cite{Guthula_Battula_Beltiukov_Guo_Gupta_2023} focused on security tasks in blockchain and network traffic.
Jiang \etal~\cite{Jiang_Wang_Liu_Xu_Tan_Zhang_2023} and Li \etal~\cite{Li_Qu_Yin_2021} developed LLMs for binary analysis.
Wang \etal~\cite{SmartInv24} introduced SmartInv for invariant inference in smart contracts.
Our work complements these studies by evaluating mainstream LLMs' reasoning in vulnerability analysis, without task-specific retraining or model customization.

%% file: conclude.tex
\section{Conclusion}
\label{sec:conclude}

This paper introduced \name, a unified and modular evaluation framework designed to decouple and enhance LLMs' inherent vulnerability reasoning.
To support fair, extensible, and reproducible evaluations, we constructed \dataset, the first benchmark that provides retrievable knowledge and context-supplementable code across three representative programming languages: Solidity, Java, and C/C++.
By applying \name to 294 vulnerable and non-vulnerable cases in 3,528 scenarios, we conducted a comprehensive study on how different enhancement strategies, namely knowledge enhancement, context supplementation, and prompt schemes, impact LLMs' performance.

%% file: appendix.tex
\appendix

\subsection{Prompts Used in \name and \dataset}
\label{sec:summarizePrompt}

\begin{tcolorbox}[title=Prompts for Summarizing the Functionalities and Root Causes from Vulnerability Reports, breakable]

    \textbf{Summarize Functionalities}\\
    Given the following vulnerability description, following the task:\\
    1. Describe the functionality implemented in the given code. This should be answered under the section "Functionality:" and written in the imperative mood, e.g., "Calculate the price of a token." Your response should be concise and limited to one paragraph and within 40-50 words.\\
    2. Remember, do not contain any variable or function or experssion name in the Functionality Result, focus on the functionality or business logic itself.\\

    \tcbline

    \textbf{Summarize Root Cause}\\
    Please provide a comprehensive and clear abstract that identifies the fundamental mechanics behind a specific vulnerability, ensuring that this knowledge can be applied universally to detect similar vulnerabilities across different scenarios. Your abstract should:\\
    1. Avoid mentioning any moderation tools or systems.\\
    2. Exclude specific code references, such as function or variable names, while providing a general yet precise technical description.\\
    3. Use the format: KeyConcept:xxxx, placing the foundational explanation of the vulnerability inside the brackets.\\
    4. Guarantee that one can understand and identify the vulnerability using only the information from the VulnerableCode and this KeyConcept.\\
    5. Strive for clarity and precision in your description, rather than brevity.\\
    6. Break down the vulnerability to its core elements, ensuring all terms are explained and there are no ambiguities.\\
    By following these guidelines, ensure that your abstract remains general and applicable to various contexts, without relying on specific code samples or detailed case-specific information.\\
    
\end{tcolorbox}

\begin{tcolorbox}[title=Prompt for Instruction Following and Auto-Annotation, breakable]
    \textbf{Generate Type and Description}\\
    I will give you some text generated by another LLM. But the format may be wrong. You must call the report API to report the result.

    \tcbline
    \textbf{Compare Types between Output and Ground Truth}\\
    You are a senior code auditor. Now I will give you a ground truth of vulnerability, and a description written by an auditor. You need to help me identify whether the description given by the auditor contains a vulnerability in the ground truth. Please report the result using the function call.\\
    Ground truth: 
    \blue{\{Ground Truth\}}\\
    Description: 
    \blue{\{Output\}}

\end{tcolorbox}

\begin{tcolorbox}[title=Prompt for Data Augmentation, breakable]

    \textbf{Vulnerable Code Generator}\\

    \blue{[\%CWE\_INFO\%]} \\

    Base on the given CWE information, please help me generate 10 different vulnerable code snippets in \blue{[\%LANGUAGE\%]} language. Each code snippet should be different from each other, and trying to cover as many as business logic as possible. You do not need to generate the description of the vulnerabilities, only the code is needed. For each code snippet, please include it in a code block, which starts with "```" and ends with "```". 

    \tcbline

    \textbf{Vulnerability Report Generation}\\

    \blue{[\%CODE\%]} \\

    The above code has \blue{[\%CWE\_TYPE\%]} vulnerability. Please help me generate a vulnerability report for it. The report should include the following sections: 1) Vulnerability Description, 2) Vulnerable Code, 3) Root Cause, 4) Impact, 5) Mitigation. Each section should be clearly labeled and contain relevant information. The report should be concise and easy to understand.
    
\end{tcolorbox}

\subsection{Detailed LLM Vulnerability Analysis Results}
\label{sec:detailedresults}

Table~\ref{tab:evaluation_all_detail_solidity} and \ref{tab:evaluation_all_detail_java_cpp} present the raw results for Solidity and Java/C/C++, respectively, under different combinations of knowledge, context, and prompt schemes, as well as the calculated precision, recall, and F1 scores.

\begin{table*}[!t]
    \caption{Raw results of TP, FP, TN, FN, and FP\textsubscript{t} along with the calculated Precision (abbreviated as ``P''), Recall (abbreviated as ``R''), and F1 score (``F1'') under different combinations of knowledge, context, prompt scheme for Solidity.}
    \vspace{-1ex}
    \label{tab:evaluation_all_detail_solidity}
    \resizebox{\textwidth}{!}{
\begin{tabular}{lll|rrrrrrrr|rrrrrrrr|}
            \hline
            \multicolumn{3}{|l|}{\multirow{2}{*}{}}                                      & \multicolumn{8}{c|}{Raw} & \multicolumn{8}{c|}{CoT}                                                                                                                                                                                                                                                                               \\ \cline{4-19}
            \multicolumn{3}{|l|}{} & \multicolumn{1}{c|}{TP} & \multicolumn{1}{c|}{FP} & \multicolumn{1}{c|}{TN} & \multicolumn{1}{c|}{FN} & \multicolumn{1}{c|}{FP\textsubscript{t}} & \multicolumn{1}{c|}{P\textsubscript{m}} & \multicolumn{1}{c|}{R\textsubscript{m}} & \multicolumn{1}{c|}{F1\textsubscript{m}} & \multicolumn{1}{c|}{TP} & \multicolumn{1}{c|}{FP} & \multicolumn{1}{c|}{TN} & \multicolumn{1}{c|}{FN} & \multicolumn{1}{c|}{FP\textsubscript{t}} & \multicolumn{1}{c|}{P\textsubscript{m}} & \multicolumn{1}{c|}{R\textsubscript{m}} & \multicolumn{1}{c|}{F1\textsubscript{m}} \\
            \hline
            \multicolumn{3}{|c|}{} & \multicolumn{15}{c}{\textbf{Solidity with data leakage mitigation}} & \\
            \hline
    \multicolumn{1}{|l|}{\multirow{6}{*}{\rotatebox{90}{GPT-4.1}}} & \multicolumn{1}{|c|}{\multirow{2}{*}{None (*3)}}       & C &  \multicolumn{1}{r|}{6} & \multicolumn{1}{r|}{57} & \multicolumn{1}{r|}{96} & \multicolumn{1}{r|}{84} & \multicolumn{1}{r|}{63} & \multicolumn{1}{r|}{4.76} & \multicolumn{1}{r|}{3.92} & \multicolumn{1}{r|}{4.30} & \multicolumn{1}{r|}{51\dataUpC} & \multicolumn{1}{r|}{75\dataDownC} & \multicolumn{1}{r|}{78\dataDownC} & \multicolumn{1}{r|}{57\dataUpC} & \multicolumn{1}{r|}{45\dataUpC} & \multicolumn{1}{r|}{29.82\dataUpC} & \multicolumn{1}{r|}{33.33\dataUpC} & \multicolumn{1}{r|}{31.48\dataUpC}\\
\hline
\end{tabular}
    }
\end{table*}

\begin{table*}[!t]
    \caption{Raw results of TP, FP, TN, FN, and FP\textsubscript{t} along with the calculated Precision (abbreviated as ``P''), Recall (abbreviated as ``R''), and F1 score (``F1'') under different combinations of knowledge, context, prompt scheme for Java and C/C++.}
    \vspace{-1ex}
    \label{tab:evaluation_all_detail_java_cpp}
    \centering
    \resizebox{\textwidth}{!}{
        \begin{tabular}{lll|rrrrrrrr|rrrrrrrr|}
            \hline
            \multicolumn{3}{|l|}{\multirow{2}{*}{}}                                      & \multicolumn{8}{c|}{Raw} & \multicolumn{8}{c|}{CoT}                                                                                                                                                                                                                                                                               \\ \cline{4-19}
            \multicolumn{3}{|l|}{} & \multicolumn{1}{c|}{TP} & \multicolumn{1}{c|}{FP} & \multicolumn{1}{c|}{TN} & \multicolumn{1}{c|}{FN} & \multicolumn{1}{c|}{FP\textsubscript{t}} & \multicolumn{1}{c|}{P} & \multicolumn{1}{c|}{R} & \multicolumn{1}{c|}{F1} & \multicolumn{1}{c|}{TP} & \multicolumn{1}{c|}{FP} & \multicolumn{1}{c|}{TN} & \multicolumn{1}{c|}{FN} & \multicolumn{1}{c|}{FP\textsubscript{t}} & \multicolumn{1}{c|}{P} & \multicolumn{1}{c|}{R} & \multicolumn{1}{c|}{F1} \\
            \hline
            \multicolumn{3}{|c|}{} & \multicolumn{15}{c}{\textbf{Java}} & \\
            \hline
            \multicolumn{1}{|l|}{\multirow{6}{*}{\rotatebox{90}{GPT-4.1}}} & \multicolumn{1}{|c|}{\multirow{2}{*}{None (*3)}}       & C &  \multicolumn{1}{r|}{15} & \multicolumn{1}{r|}{45} & \multicolumn{1}{r|}{93} & \multicolumn{1}{r|}{93} & \multicolumn{1}{r|}{30} & \multicolumn{1}{r|}{16.67} & \multicolumn{1}{r|}{10.87} & \multicolumn{1}{r|}{13.16} & \multicolumn{1}{r|}{15} & \multicolumn{1}{r|}{33\dataUpC} & \multicolumn{1}{r|}{105\dataUpC} & \multicolumn{1}{r|}{102\dataDownC} & \multicolumn{1}{r|}{21\dataUpC} & \multicolumn{1}{r|}{21.74\dataUpC} & \multicolumn{1}{r|}{10.87} & \multicolumn{1}{r|}{14.49\dataUpC}\\
\hline
\end{tabular}
    }
\end{table*}